\documentclass[11pt]{article}
\usepackage[latin9]{inputenc}
\usepackage{amsmath}
\usepackage{amssymb}
\usepackage{graphicx}
\usepackage{esint}
\usepackage[unicode=true,
 bookmarks=false,
 breaklinks=false,pdfborder={0 0 1},backref=section,colorlinks=false]
 {hyperref}
\usepackage{breakurl}

\makeatletter
\@ifundefined{date}{}{\date{}}

\usepackage{esint}
\usepackage{latexsym}
\usepackage{graphicx}\usepackage{bm}\usepackage{longtable}

\usepackage{xcolor}

\@addtoreset{equation}{section}

\makeatother

\begin{document}

\title{Dynamically flavored description of holographic QCD in the presence
of a magnetic field}

\maketitle
\begin{center}
Si-wen Li\footnote{Email: cloudk@mail.ustc.edu.cn} and Tuo Jia\footnote{Email: jt2011@mail.ustc.edu.cn}
\par\end{center}

\begin{center}
\emph{Department of Modern Physics, }\\
\emph{ University of Science and Technology of China, }\\
\emph{ Hefei 230026, Anhui, China}
\par\end{center}

\vspace{10mm}

\begin{abstract}
We construct the gravitational solution of the Witten-Sakai-Sugimoto
model by introducing a magnetic field on the flavor brane. With taking
into account their backreaction, we re-solve the type IIA supergravity
in the presence of the magnetic field. Our calculation shows the gravitational
solutions are magnetic-dependent and analytic both in the bubble (confined)
and black brane (deconfined) case. We study the dual field theory
at the leading order in the ratio of the number of flavors and colors,
also in the Veneziano limit. Some physical properties related to the
hadronic physics in an external magnetic field are discussed by using
our confined backreaction solution holographically. We also investigate
the thermodynamics and holographic renormalization of this model in
both phases by our solution. Since the backreaction of the magnetic
field is considered in our gravitational solution, it allows us to
study the Hawking-Page transition with flavors and colors of this
model in the presence of the magnetic field. Finally we therefore
obtain the holographic phase diagram with the contributions from the
flavors and the magnetic field. Our holographic phase diagram is in
agreement with lattice QCD result qualitatively, which thus can be
interpreted as the inhibition of confinement or chirally broken symmetry
by the magnetic field.
\end{abstract}

\title{\newpage{}}

\tableofcontents{}

\title{\newpage{}}

\section{Introduction}

\subsection*{Motivation}

Recent years, in some results from lattice QCD \cite{key-1,key-2},
it seems the QCD phase could be changed by a strong magnetic field.
By the analysis of some thermodynamic observables, it has been found
the critical temperature of the crossover region should fall when
the magnetic field increases \cite{key-1,key-2}. It implies the confinement/deconfinement
phase transition or the chiral phase transition \cite{key-3} would
tend to be induced by a strong magnetic field. With the MIT bag model,
this result could be reproduced qualitatively \cite{key-4}, reflecting
the great significance of quark confinement. Furthermore, the approach
of large-$N_{c}$ QCD has already been considered in \cite{key-5}.
From the analysis of the flavor correction $N_{f}/N_{c}$ to the pressure,
this effect has also been obtained due to the quark degrees of freedom.
On the other hand, gauge/gravity duality or AdS/CFT has become a framework
to understand non-perturbative aspects of strong-coupled quantum field
theory \cite{key-6,key-7,key-8}. Therefore, the motivation of our
work is to investigate the thermodynamics of the quarks and gluons
by using the holographic method in the presence of a magnetic field
with considering the dynamics of the flavors\footnote{To compare our results with lattice QCD, we will discuss the case
with zero chemical potential throughout our manuscript.}.

\subsection*{Model }

The famous Witten-Sakai-Sugimoto model \cite{key-9,key-10} is the
model which currently becomes closest to QCD since it has been proposed
to holographically study the non-perturbative QCD for a long time,
for examples \cite{key-11,key-12,key-13,key-14,key-15,key-16,key-17,key-18,key-19,key-20,key-21}.
By the underlying string theory, this model describes a non-supersymmetric
and non-conformal Yang-Mills theory in 3+1 dimension coupled to $N_{f}$
chiral massless fermions (quarks) and adjoint massive matter, as a
low energy effective theory. In this model, there are $N_{c}$ D4-branes
compactified on a circle representing the dynamics of gluons, $N_{f}$
species of massless quarks introduced by putting in $N_{f}$ pairs
of D8 and anti D8-branes ($\overline{\mathrm{D8}}$-branes). By taking
the large $N_{c}$ limit i.e. $N_{f}\ll N_{c}\rightarrow\infty$,
these $N_{c}$ D4-branes produce a 10D background geometry described
by type IIA supergravity while the $N_{f}$ $\mathrm{D}8/\overline{\mathrm{D}8}$-branes
are as probes. Accordingly, the fundamental quarks do not have dynamical
degrees of freedoms thus they are quenched. 

Furthermore, the description of deconfinement transition and chiral
transition in the Witten-Sakai-Sugimoto model was proposed in \cite{key-11}.
At zero temperature, the bubble (confined) solution of the $N_{c}$
D4-branes is dominant, corresponding the confinement phase in the
dual field theory, while the black brane (deconfined) solution of
the $N_{c}$ D4-branes arises as the deconfinement phase at high temperature.
Thus the phase transition between confinement and deconfinement can
be identified as the Hawking-Page transition between two different
background geometries. We can therefore evaluate the critical temperature
by the analysis of the pressure in bubble and black brane background.
The result shows that a confinement/deconfinement transition at $T_{c}=M_{KK}/2\pi$
arises, where $M_{KK}$ is a mass scale of the mass spectrum. However,
while the bubble solution can be connected to the confinement phase
of the dual field theory, this is less clear for the black brane solution
because of the mismatched value of the Polyakov loops\footnote{This is the reason we use ``confined/deconfined geometry'' instead
of ``confinement/deconfinement''. In fact, the behavior of the Witten-Sakai-Sugimoto
model interpolates between NJL and QCD according to \cite{key-15,key-30,key-31,key-32,key-33}.} \cite{key-17,key-22,key-23}, which thus makes ``the black D4-brane
solution corresponding the deconfinement phase'' may not be strictly
rigorous. Nevertheless, we can focus on the chiral transition in this
setup since the embedded flavor branes take connected/parallel configuration
in the bubble/black D4-brane solution respectively, which corresponds
to the chirally broken/symmetric phase in the dual field theory.

\subsection*{Goal and method}

Our goals for this paper are collected as follows,
\begin{enumerate}
\item Construct gravitational solutions of this model by taking account
of the backreaction from the flavor brane with a magnetic field. Then
investigate some physical quantities in confinement phase with our
solution.
\item Study the holographic renormalization, thermodynamics and Hawking-Page
transition with our gravitational solutions. Compare our results with
lattice QCD \cite{key-1,key-2}.
\end{enumerate}
In order to achieve the goals above, let us outline some technical
details in our manuscript. First, we use the smearing technique \cite{key-24,key-25,key-26,key-27,key-28}
for the flavor branes to construct gravitational solutions as \cite{key-18}
in the presence of a magnetic field (with zero chemical potential).
Then in order to preserve the isometries of the original background,
we also homogeneously smear a large number $N_{f}$ of D8-branes on
the $x_{4}$ circle where the $N_{c}$ D4-branes are wrapped. As it
will be seen, while this configuration simplifies the calculations
greatly, we have to solve a set of coupled second order equations
of motion of this system. Because of the presence of the magnetic
field, these equations of motion are all highly non-linear which are
still extremely complicated to solve. Hence we focus on solving these
equations in the limit of small magnetic field and small flavor backreaction
since it admits analytically magnetic-dependent solutions. To determine
the integration constants in our solution, we furthermore require
that the backgrounds must be completely regular in the IR region of
the dual field theory. With the presence of the magnetic field, the
integration constants could be able to depend on the constant magnetic
field. However, we find it is not enough to determine all the integration
constants just by these geometric requirements. Besides, in the UV
region, there also is an non-removable divergence unaffected by the
presence of the magnetic field, which is due to the Landau pole in
field theory, reflecting in the running coupling holographically.

Last but not least, since the onshell action evaluated by our gravitational
solutions is divergent, we need to holographically renormalize the
theory in order to study its thermodynamics. The counterterms have
been computed with a magnetic field and we find if the parameters
in the covariant counterterms depend on the magnetic field, they are
enough to cancel all the divergences in our calculations. Then we
can obtain the phase diagram by comparing the renormalized confined/deconfined
pressure. Our holographic phase diagram shows, the critical temperature
decreases when the magnetic field increases in the probe approximation,
which qualitatively agrees with the lattice QCD results \cite{key-1,key-2}.

\subsection*{Relation to previous works and outline}

The thermodynamics of holographic QCD with this model has been widely
studied in many present works \cite{key-11,key-15,key-16,key-17,key-19,key-20,key-21,key-29},
however the backreaction case is not considered in these works. Particularly,
in \cite{key-18}, the dynamical flavors have been taken into account
in this model without magnetic field. In \cite{key-29}, the thermodynamics
of the quarks and gluons in the presence of a magnetic field has been
studied (without the flavored backreaction). Thus we would like to
combine \cite{key-18} with \cite{key-29} in this manuscript. Technically,
our calculation is an extension of \cite{key-18} by introducing the
dependence of the magnetic field, so we will employ the similar conventions
as in \cite{key-18}.

This paper is organized as follows. In the next Section 2, we will
give a brief review of the Witten-Sakai-Sugimoto model. In Section
3, we introduce a magnetic field on the flavor brane and construct
the gravitational solution by re-solving the type IIA plus flavor
brane action, both in confined (bubble) case and deconfined (black
brane) case. The magnetic-dependent solution is also given in this
section. In Section 4, we discuss some physical quantities by imposing
the constructed solution with some special constraints in the confined
case. In section 5, it shows the holographic renormalization in our
calculation, then we evaluate the renormalized onshell action and
the counterterms by our magnetic-dependent solutions. In Section 6,
we discuss the holographic phase diagram with the magnetic field in
the case of the probe approximation and backreaction respectively,
then compare our results with lattice QCD. Discussion and summary
are given in the final section.

\section{Reviews of the Witten-Sakai-Sugimoto model}

In this section, we review the Witten-Sakai-Sugimoto model systematically.

A non-supersymmetric and non-conformal (3+1 dimensional) Yang-Mills
theory was proposed by Witten \cite{key-34} as the low energy limit
of a Kaluza-Klein (KK) reduction of a 5+1 dimensional $SU\left(N_{c}\right)$
super conformal theory which couples to massless adjoint scalar and
fermions. This theory is the low energy effective theory describing
the open string ending on the worldvolume of $N_{c}$ coincident D4-branes
placed in the 10D Minkowskian spacetime. By the dimensional reduction,
the theory is compactified on a circle (denoted as $x_{4}$) of length
$\beta_{4}$. With the choice of boundary conditions for bosons (periodic
b.c.) and fermions (anti-periodic b.c.), the massless modes at low
energy scale i.e. $E\ll1/\beta_{4}$ are the gauge fields of 3+1 dimensional
$SU(N_{c})$ Yang-Mills theory. The supersymmetry breaks down since
the other modes (including fermions) get masses $M_{kk}\sim1/\beta_{4}$.
If $T_{s}/M_{KK}=2\lambda_{4}/27\pi\ll1$, where $T_{s},\ \lambda_{4}$
is the string tension and 4d 't Hooft coupling respectively, the low
energy theory could be decoupled from the Kaluza-Klein modes.

However, as it is known there is not any simple description in the
most interesting region $\lambda_{4}\sim1$ in Witten's model. As
a conjecture by holography, there should be a dual description in
terms of a classical gravity theory on a background arising as the
near-horizon limit of sourced $N_{c}$ D4-branes, we can therefore
obtain many detailed informations in the region of $\lambda_{4}\gg1$.
Such a background produced by $N_{c}$ D4-branes would have the topology
of a product $\mathbb{R}^{1,3}\times\mathbb{R}_{u}\times S_{x_{4}}\times S^{4}$.
Here $\mathbb{R}^{1,3}$ represents the 3+1 dimensional spacetime
where we live in. $\mathbb{R}_{u}$ represents the radial direction
denoted by the coordinate $u$ as the holographic direction, which
could be roughly treated as the energy scale of the renormalization
group in the dual field theory. In the $\left(u,\ x_{4}\right)$ plane
of the subspace, the confined background looks like a cigar and the
size of the $x_{4}$ circle smoothly shrinks to zero at a finite value
$u_{KK}$ of the radial coordinate $u$. $S^{4}$ represents the additional
dimensions, whose isometry group is $SO\left(5\right)$ identified
as a global symmetry group under rotation of the massive Kaluza-Klein
fields. The theory describes confinement in the dual field theory
and the chiral symmetry breaks at zero temperature once it couples
to the chiral massless quarks.

It is achieved to add a stack of $N_{f}$ pairs of suitably $\mathrm{D}8/\overline{\mathrm{D}8}$-branes
embedded in the $N_{c}$ D4-branes background geometry to introduce
$N_{f}$ chiral fundamental massless quarks as \cite{key-9} in Witten's
model. Quarks are in the fundamental representation of color and flavor
group since they come from the massless spectrum of the open strings
which are stretching between the color and flavor branes. Because
the flavor $\mathrm{D}8/\overline{\mathrm{D}8}$-branes are probes
in this system, their backreaction to the geometric background is
neglected. Correspondingly, the fundamental quarks in the dual field
theory are in the quenched approximation. Besides, the flavor branes
offer a $U_{R}\left(N_{f}\right)\times U_{L}\left(N_{f}\right)$ symmetry
which could be identified as the global flavor symmetry holographically.
Then it is recognized that the flavor branes connect to each other
as a U-shape at zero temperature representing chirally broken symmetry
automatically.

In the bubble (confined) background, the geometry is described by
the bubble solution of $N_{c}$ D4-brane with the following metric,

\begin{align}
ds^{2} & =\ \left(\frac{u}{R}\right)^{3/2}\left[\eta_{\mu\nu}dx^{\mu}dx^{\nu}+f\left(u\right)dx_{4}^{2}\right]+\left(\frac{R}{u}\right)^{3/2}\left[\frac{du^{2}}{f\left(u\right)}+u^{2}d\Omega_{4}^{2}\right]\nonumber \\
e^{\phi} & =\ g_{s}\left(\frac{u}{R}\right)^{3/4},\ \ \ F_{4}=\frac{3N_{c}}{4\pi}\omega_{4},\ \ \ f\left(u\right)=1-\frac{u_{KK}^{3}}{u^{3}}.\label{eq:1}
\end{align}
where $R^{3}=\pi g_{s}N_{c}l_{s}^{3}$ is the curvature radius of
the background geometry and $\omega_{4}$ is the volume form of $S^{4}$,
$g_{s}$ and $l_{s}$ is the string coupling and length respectively.
$\phi$ is dilaton and $F_{4}$ is the Ramond-Ramond four form. For
the index, we have defined $\mu=0,1,2,3$. At the scale $M_{KK}$,
the 't Hooft coupling is defined as $\lambda_{4}=g_{YM}^{2}N_{c}=4\pi^{2}g_{s}N_{c}l_{s}/\beta_{4}$
in the 4-dimensional theory. Since $f\left(u_{KK}\right)=0$, the
$x_{4}$ circle shrinks at $u=u_{KK}$. In order to omit the conical
singularities at $u=u_{KK}$, it provides the following relation,

\begin{equation}
9\beta_{4}^{2}u_{KK}=16\pi^{2}R^{3}.
\end{equation}
Here $\beta_{4}$, as the length of the $x_{4}$ circle, is related
to the mass scale $M_{KK}$ by $\beta_{4}=2\pi/M_{KK}$. 

There is an alternatively allowed solution which is the black brane
(deconfined) solution taking the following metric,

\begin{align}
ds^{2} & =\ \left(\frac{u}{R}\right)^{3/2}\left[-f_{T}\left(u\right)dt^{2}+\delta_{ij}dx^{i}dx^{j}+dx_{4}^{2}\right]+\left(\frac{R}{u}\right)^{3/2}\left[\frac{du^{2}}{f_{T}\left(u\right)}+u^{2}d\Omega_{4}^{2}\right]\nonumber \\
f_{T}\left(u\right) & =\ 1-\frac{u_{T}^{3}}{u^{3}}.\label{eq:3}
\end{align}
Here $i,j=1,2,3$. Similarly, it provides the following relation with
the $x_{0}$ circle smoothly shrinking to zero at the horizon $u=u_{T}$, 

\begin{equation}
9\beta^{2}u_{T}=16\pi^{2}R^{3}.
\end{equation}
Therefore, we have the Hawking temperature as

\begin{equation}
T=1/\beta,
\end{equation}
where $\beta$ is the length of $x^{0}$ in the deconfined geometry.

\begin{figure}
\begin{centering}
\includegraphics[scale=0.15]{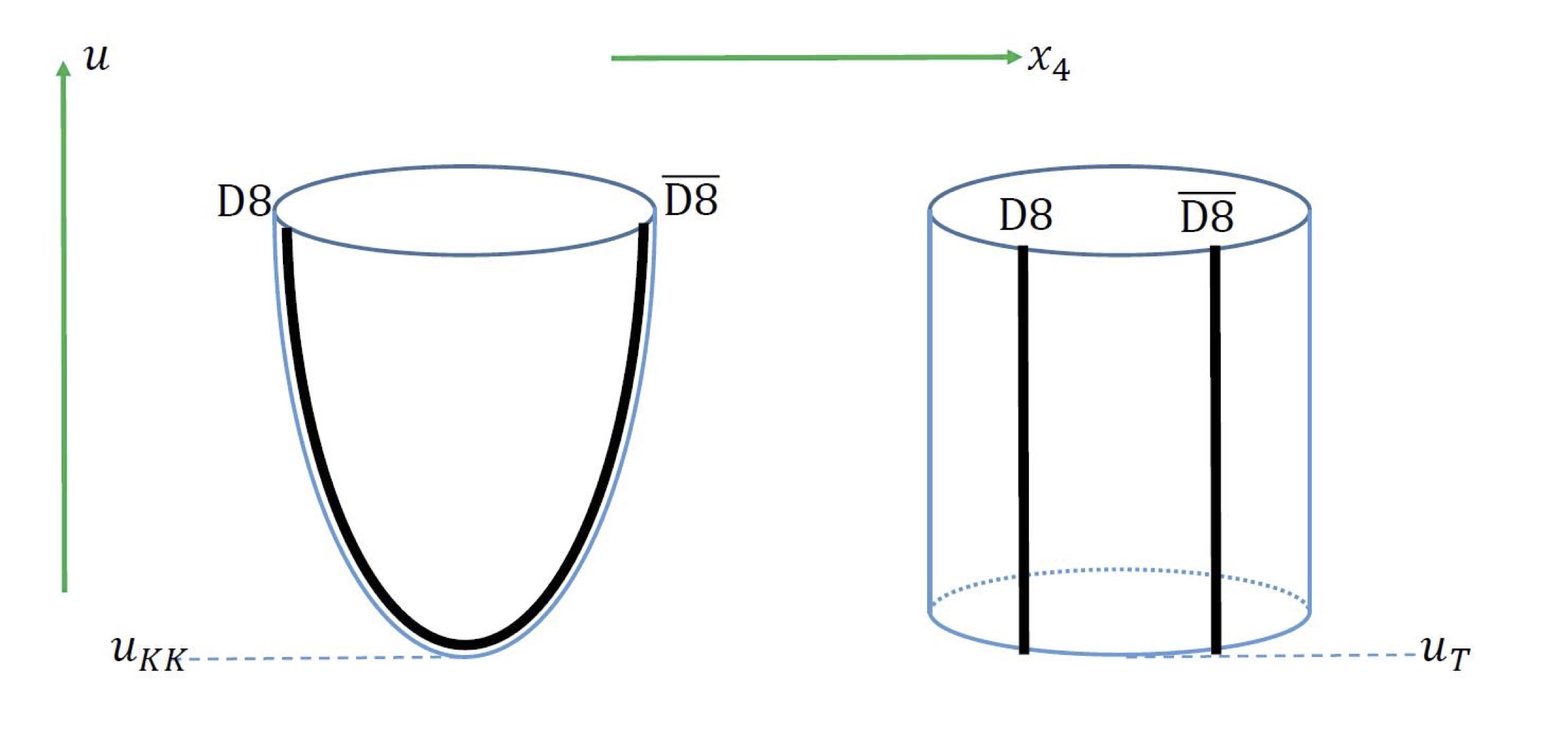}
\par\end{centering}

\caption{Phase Transition in our model.\textbf{ Left}: In confined geometry
the flavor $\mathrm{D}8/\overline{\mathrm{D}8}$-branes are located
at antipodal points of $x_{4}$ - circle and always connected, realizing
as chirally broken phase. \textbf{Right}: In deconfined geometry the
flavor $\mathrm{D}8/\overline{\mathrm{D}8}$-branes could be parallel,
realizing as chirally symmetric phase.}
\end{figure}

In the classical limit, the gravity partition function $Z\sim e^{-S}$,
which is related to the Euclidean onshell action, could be identified
as the free energy of the system thermodynamically. The phase diagram
can be obtained by comparing the free energy of the two phases above.
It has turned out the bubble solution is dominant at zero temperature,
while the black brane solution of the $N_{c}$ D4-branes arises at
high temperature, which provides the critical temperature of the phase
transition as,

\begin{equation}
T_{c}=1/\beta_{4}=\frac{M_{KK}}{2\pi}.
\end{equation}

In this manuscript, we are going to work in the following configuration
when the flavor branes are considered. That is, in the confined case,
the $\mathrm{D}8/\overline{\mathrm{D}8}$-branes are placed at antipodal
points of $x_{4}$ - circle. When the temperature increases, the connected
position on $\mathrm{D}8/\overline{\mathrm{D}8}$-branes falls into
the horizon in the $T>T_{c}$ phase as in Figure 1. It thus turns
to the deconfined case and the $\mathrm{D}8/\overline{\mathrm{D}8}$-branes
become parallel (disconnected). Accordingly, the chiral symmetry is
restored since the flavor symmetry group remains $U_{R}\left(N_{f}\right)\times U_{L}\left(N_{f}\right)$
in the configuration of a stake of parallel $\mathrm{D}8/\overline{\mathrm{D}8}$-branes.
So the confined/deconfined or chiral phase transition could be identified
as the Hawking-Page transition of the background with the connected/parallel
configuration of the flavor branes.

\section{Solutions with the flavored backreaction in the presence of a magnetic
field}

In the following sections, there would be three relevant and useful
coordinates which are $\rho,\ x,\ r$. For the reader convenience,
the relation of these coordinates and the standard $u$ coordinate
used in the Witten-Sakai-Sugimoto model (\ref{eq:1}) (\ref{eq:3})
is summarized as follows,

\begin{equation}
e^{-3r}=1-\frac{u_{0}^{3}}{u^{3}},\ r=a\rho,\ x=e^{-3r/2}.\label{eq:7}
\end{equation}

\noindent As it will be seen that $u_{0}$ represents $u_{KK}$ in
confined geometry or $u_{T}$ in deconfined geometry. We are going
to use $a_{T}$ to replace $a$ in (\ref{eq:7}) in the deconfined
geometry and the explicit definition of $a$ or $a_{T}$ could be
found in the following relevant formulas (in Eq.(\ref{eq:19}) and
Eq.(\ref{eq:38})). Since our calculation is an extension of \cite{key-18},
we will employ the similar conventions as in \cite{key-18}.

\subsection{Confined geometry}

\subsubsection*{Ansatz and solution}

In the Witten-Sakai-Sugimoto model \cite{key-9}, the flavor $\mathrm{D}8/\overline{\mathrm{D}8}$-branes
are treated as the probes embedded in the confined geometry. However,
in this subsection we would like to take into account their backreaction
to the first order of $N_{f}/N_{c}$ in the confined case. Hence we
are going to use the same trick as \cite{key-18}, that is to consider
a setup where $N_{f}$ D8-branes are smeared homogeneously along the
transverse $x_{4}$ circle \cite{key-24,key-25,key-26,key-27,key-28}.
And we will consider the model below the critical temperature ($T<T_{c}$)
with a background magnetic field on the flavor branes. 

For $T=0$, the ansatz of the metric in string frame is given as \cite{key-11},

\begin{equation}
ds^{2}=e^{2\lambda}\left(-dt^{2}+dx_{a}dx^{a}\right)+e^{2\tilde{\lambda}}dx_{4}^{2}+l_{s}^{2}e^{-2\varphi}d\rho^{2}+l_{s}^{2}e^{2\nu}d\Omega_{4}^{2},\label{eq:8}
\end{equation}

\noindent where $a=1,2,3$ and $\lambda,\tilde{\lambda},\nu,\phi$
are functions depended on the holographic coordinate $\rho$ only.
$x_{4}$ is the compactified coordinate on a circle with the length
$\beta_{4}=2\pi/M_{KK}$. The function $\varphi$ are defined as

\begin{equation}
\varphi=2\phi-4\lambda-\tilde{\lambda}-4\nu.\label{eq:9}
\end{equation}

\noindent In order to take into account the backreaction of the flavor
and the magnetic field, we have to consider the total action in type
IIA supergravity with the presence of a magnetic field on the flavor
branes. The relevant action (bulk fields plus smeared flavor brane)
is,

\begin{align}
S & =\frac{1}{2k_{0}^{2}}\int d^{10}x\sqrt{-g}\left[e^{-2\phi}\left(\mathcal{R}+4\left(\partial\phi\right)^{2}\right)-\frac{1}{2}\left|F_{4}\right|^{2}\right]\nonumber \\
 & \ \ -\frac{N_{f}T_{8}M_{KK}}{\pi}\int d^{10}x\frac{\sqrt{-\det\left(g+2\pi\alpha^{\prime}F\right)}}{\sqrt{g_{44}}}e^{-\phi}.\label{eq:10}
\end{align}

\noindent The first part of (\ref{eq:10}) is the action of the bulk
fields while the last part arises as the contribution from the Dirac-Born-Infield
(DBI) action of $N_{f}$ D8-branes which are smeared on the transverse
$x_{4}$ circle. Here $k_{0}$ is related to the 10d Newton coupling.
In confined geometry, we consider the antipodal configuration for
the flavor branes and put the smeared DBI action on-shell i.e. the
embedding coordinate $x_{4}=x_{4}\left(\rho\right)$ satisfies its
equation of motion $\frac{d}{d\rho}x_{4}=0$. The integration over
the radial coordinate has been calculated as two times to account
for the presence of two branches at two antipodal points on the $x_{4}$.
Furthermore, we have turned on a $U\left(1\right)$ gauge field on
the flavor branes which is the dual of an external background magnetic
field. Thus as \cite{key-19,key-29,key-30} we set a constant magnetic
field $2\pi\alpha^{\prime}F_{12}=b$, where $b$ is dimensionless
constant\footnote{Notice that the Wess-Zumino term of the $\mathrm{D}8/\overline{\mathrm{D}8}$-brane
action vanishes since only one component of the gauge field strength
is turned on. And as a consistent solution for the DBI action, it
is allowed to set the magnetic field as a constant. See also \cite{key-19,key-29,key-30,key-35,key-36,key-37}
for the similar setup.}. With the implementation of the ansatz (\ref{eq:8}), it yields the
following 1d action \cite{key-11,key-18},

\begin{eqnarray}
S & = & \mathcal{V}\int d\rho\left[-4\dot{\lambda}^{2}-\dot{\widetilde{\lambda}}^{2}-4\dot{\nu}^{2}+\dot{\varphi}^{2}+V+\mathrm{total\ derivatives}\right],\nonumber \\
V & = & 12e^{-2\nu-2\varphi}-Q_{c}^{2}e^{4\lambda+\widetilde{\lambda}-4\nu-\varphi}-Q_{f}e^{2\lambda-\frac{\widetilde{\lambda}}{2}+2\nu-\frac{3}{2}\varphi}\sqrt{1+b^{2}e^{-4\lambda}},\label{eq:11}
\end{eqnarray}

\noindent where we have defined

\begin{align}
R^{3} & =\pi g_{s}N_{c}l_{s}^{3},\ \ Q_{c}=\frac{3}{\sqrt{2}g_{s}}\frac{R^{3}}{l_{s}^{3}}=\frac{3\pi N_{c}}{\sqrt{2}},\ \ \epsilon_{f}=\frac{R^{3/2}u_{0}^{1/2}g_{s}}{l_{s}^{2}}Q_{f},\nonumber \\
Q_{f} & =\frac{2k_{0}^{2}N_{f}T_{8}M_{KK}l_{s}^{2}}{\pi},\ \ \mathcal{V}=\frac{1}{2k_{0}^{2}}V_{3}V_{S^{4}}\frac{1}{T}\frac{2\pi}{M_{KK}}l_{s}^{3},
\end{align}

\noindent Note that we are going to use parameter $\epsilon_{f}$
(or $\epsilon_{fT}$, in the deconfined case) to weigh the contribution
from flavors to the action, and the dot represents the derivatives
are w.r.t. $\rho$. Moreover action (\ref{eq:11}) has to be supported
by the zero-energy constraint \cite{key-18,key-38},

\begin{equation}
-4\dot{\lambda}^{2}-\dot{\widetilde{\lambda}}^{2}-4\dot{\nu}^{2}+\dot{\varphi}^{2}=V,\label{eq:13}
\end{equation}
which makes the equations of motion from 10d action (\ref{eq:10})
and the effective 1d action (\ref{eq:11}) coincident if the homogeneous
ansatz (\ref{eq:8}) is adopted. Then the equations of motion from
the previous action (\ref{eq:11}) are as follows (derivatives are
w.r.t. $\rho$),

\begin{eqnarray}
\ddot{\lambda}-\frac{1}{2}Q_{c}^{2}e^{8\lambda+2\widetilde{\lambda}-2\phi} & = & \frac{1}{4}Q_{f}e^{8\lambda+8\nu-3\phi+\widetilde{\lambda}}\frac{1}{\sqrt{1+b^{2}e^{-4\lambda}}},\nonumber \\
\ddot{\widetilde{\lambda}}-\frac{1}{2}Q_{c}^{2}e^{8\lambda+2\widetilde{\lambda}-2\phi} & = & -\frac{1}{4}Q_{f}e^{8\lambda+8\nu-3\phi+\widetilde{\lambda}}\sqrt{1+b^{2}e^{-4\lambda}},\nonumber \\
\ddot{\phi}-\frac{1}{2}Q_{c}^{2}e^{8\lambda+2\widetilde{\lambda}-2\phi} & = & \frac{Q_{f}e^{8\lambda+8\nu-3\phi+\widetilde{\lambda}}\left(3b^{2}e^{-4\lambda}+5\right)}{4\sqrt{1+b^{2}e^{-4\lambda}}},\nonumber \\
\ddot{\nu}-3e^{6\nu-4\phi+8\lambda+2\widetilde{\lambda}}+\frac{1}{2}Q_{c}^{2}e^{8\lambda+2\widetilde{\lambda}-2\phi} & = & \frac{1}{4}Q_{f}e^{8\lambda+8\nu-3\phi+\widetilde{\lambda}}\sqrt{1+b^{2}e^{-4\lambda}}.\label{eq:14}
\end{eqnarray}

\noindent We have used the definition of (\ref{eq:9}) to replace
$\varphi$ by the dilaton field $\phi$. However, we will not attempt
to solve equations (\ref{eq:14}) exactly, instead, we will focus
on the small magnetic field case i.e. keeping only the leading $b^{2}$
term. On the other hand, since our concern is to find a perturbative
solution of (\ref{eq:14}) at the first order of the parameter $\epsilon_{f}$,
we write all the relevant functions in (\ref{eq:14}) as,

\begin{equation}
\Psi\left(r\right)=\Psi_{0}\left(r\right)+\epsilon_{f}\Psi_{1}\left(r\right)+\mathcal{O}\left(\epsilon_{f}^{2}\right),\label{eq:15}
\end{equation}

\noindent Then we use the following unflavored solutions as the zeroth
order solution\footnote{Functions (\ref{eq:16}) and (\ref{eq:17}) are nothing but the compacted
D4-brane solution used in the Witten-Sakai-Sugimoto model, expressed
in $r$ coordinate with the coordinate transformation (\ref{eq:7}).}, 

\begin{eqnarray}
\lambda_{0}\left(r\right) & = & f_{0}\left(r\right)+\frac{3}{4}\log\frac{u_{0}}{R},\nonumber \\
\widetilde{\lambda}_{0}\left(r\right) & = & f_{0}\left(r\right)-\frac{3}{2}r+\frac{3}{4}\log\frac{u_{0}}{R},\nonumber \\
\phi_{0}\left(r\right) & = & f_{0}\left(r\right)+\frac{3}{4}\log\frac{u_{0}}{R}+\log g_{s},\nonumber \\
\nu_{0}\left(r\right) & = & \frac{1}{3}f_{0}\left(r\right)+\frac{1}{4}\log\frac{u_{0}}{R}+\log\frac{R}{l_{s}},\label{eq:16}
\end{eqnarray}

\noindent with

\begin{equation}
f_{0}\left(r\right)=-\frac{1}{4}\log\left(1-e^{-3r}\right).\label{eq:17}
\end{equation}

\noindent In order to keep the leading $b^{2}$ terms, we have the
following equations from (\ref{eq:14}) for the leading order function
$\Psi_{1}\left(r\right)$ in the expansion of (\ref{eq:15}), (derivatives
are w.r.t. $r$),

\begin{eqnarray}
\lambda_{1}^{\prime\prime}-\frac{9}{2}\frac{e^{-3r}}{\left(1-e^{-3r}\right)^{2}}\left(4\lambda_{1}+\widetilde{\lambda}_{1}-\phi_{1}\right) & = & \frac{1}{4}\frac{e^{-3r/2}}{\left(1-e^{-3r}\right)^{13/6}}\left[1-\frac{1}{2}q_{b}^{2}\left(1-e^{-3r}\right)\right],\nonumber \\
\widetilde{\lambda}_{1}^{\prime\prime}-\frac{9}{2}\frac{e^{-3r}}{\left(1-e^{-3r}\right)^{2}}\left(4\lambda_{1}+\widetilde{\lambda}_{1}-\phi_{1}\right) & = & -\frac{1}{4}\frac{e^{-3r/2}}{\left(1-e^{-3r}\right)^{13/6}}\left[1+\frac{1}{2}q_{b}^{2}\left(1-e^{-3r}\right)\right],\nonumber \\
\phi_{1}^{\prime\prime}-\frac{9}{2}\frac{e^{-3r}}{\left(1-e^{-3r}\right)^{2}}\left(4\lambda_{1}+\widetilde{\lambda}_{1}-\phi_{1}\right) & = & \frac{1}{4}\frac{e^{-3r/2}}{\left(1-e^{-3r}\right)^{13/6}}\left[5+\frac{1}{2}q_{b}^{2}\left(1-e^{-3r}\right)\right],\nonumber \\
\nu_{1}^{\prime\prime}-\frac{3}{2}\frac{e^{-3r}}{\left(1-e^{-3r}\right)^{2}}\left(12\nu_{1}+4\lambda_{1}-5\phi_{1}+\widetilde{\lambda}_{1}\right) & = & \frac{1}{4}\frac{e^{-3r/2}}{\left(1-e^{-3r}\right)^{13/6}}\left[1+\frac{1}{2}q_{b}^{2}\left(1-e^{-3r}\right)\right].\nonumber \\
\label{eq:18}
\end{eqnarray}
Here $q_{b}=bR^{3/2}u_{0}^{-3/2}$ and we have assumed $\epsilon_{f}=\frac{R^{3/2}u_{0}^{1/2}g_{s}}{l_{s}^{2}}Q_{f}=\frac{1}{12\pi^{3}}\lambda_{4}^{2}\frac{N_{f}}{N_{c}}\ll1$.
$\lambda_{4}$ is the 't Hooft coupling constant which should be fixed.
Other relevant parameters are defined as

\begin{equation}
r=a\rho,\ a=\frac{\sqrt{2}Q_{c}u_{0}^{3}}{3R^{3}g_{s}}=\frac{u_{0}^{3}}{l_{s}^{3}g_{s}^{2}},\ \lambda_{4}=g_{YM}^{2}N_{c}.\label{eq:19}
\end{equation}

\noindent With the equations in (\ref{eq:18}), we find that,

\begin{eqnarray}
\widetilde{\lambda}_{1} & = & \lambda_{1}-\frac{1}{2}f-A_{1}-B_{1}r,\nonumber \\
\phi_{1} & = & \lambda_{1}+f+\frac{q_{b}^{2}}{4}h-A_{2}-B_{2}r,\label{eq:20}
\end{eqnarray}

\noindent where $A_{1,2}$ and $B_{1,2}$ are integration constants
and $f,\ h$ are two particular functions which satisfy

\begin{eqnarray}
f^{\prime\prime}\left(r\right) & = & \frac{e^{-3r/2}}{\left(1-e^{-3r}\right)^{13/6}},\nonumber \\
h^{\prime\prime}\left(r\right) & = & \frac{e^{-3r/2}}{\left(1-e^{-3r}\right)^{7/6}}.
\end{eqnarray}

\noindent The equations in (\ref{eq:18}) would be quite easy to solve
after a re-combination and the definition of $4y=4\lambda_{1}+\widetilde{\lambda}_{1}-\phi_{1}$,
it yields an equation for $y$ which is\footnote{Similarly, we also find an equation for the function $v$ which is
used in (\ref{eq:23}).},

\begin{equation}
y^{\prime\prime}-\frac{18e^{-3r}}{\left(1-e^{-3r}\right)^{2}}y+\frac{1}{8}\frac{e^{-3r/2}}{\left(1-e^{-3r}\right)^{13/6}}+\frac{3}{16}q_{b}^{2}\frac{e^{-3r/2}}{\left(1-e^{-3r}\right)^{5/6}}=0.\label{eq:22}
\end{equation}

\noindent So we have the following solution expressed in terms of
generalized hypergeometric functions as\footnote{As a quick check, our solution will return to \cite{key-18} once
we turn off the magnetic field.},

\begin{eqnarray}
\lambda_{1} & = & \frac{q_{b}^{2}}{16}h+\frac{3}{8}f+y-\frac{1}{4}\left(A_{2}-A_{1}\right)-\frac{1}{4}\left(B_{2}-B_{1}\right)r,\nonumber \\
\widetilde{\lambda}_{1} & = & \frac{q_{b}^{2}}{16}h-\frac{1}{8}f+y-\frac{1}{4}\left(A_{2}+B_{2}r\right)-\frac{3}{4}\left(A_{1}+B_{1}r\right),\nonumber \\
\phi_{1} & = & \frac{5q_{b}^{2}}{16}h+\frac{11}{8}f+y+\frac{1}{4}\left(A_{1}+B_{1}r\right)-\frac{5}{4}\left(A_{2}+B_{2}r\right),\nonumber \\
\nu_{1} & = & \frac{5q_{b}^{2}}{48}h+\frac{11}{24}f+\frac{1}{3}\left(y-v\right)+\frac{1}{12}\left(A_{1}+B_{1}r\right)-\frac{5}{12}\left(A_{2}+B_{2}r\right).\label{eq:23}
\end{eqnarray}

\noindent And the relevant functions in (\ref{eq:23}) are

\begin{eqnarray}
f & = & \frac{4}{9}e^{-3r/2}\ _{3}F_{2}\left(\frac{1}{2},\frac{1}{2},\frac{13}{6};\frac{3}{2},\frac{3}{2};e^{-3r}\right),\nonumber \\
h & = & \frac{4}{9}e^{-3r/2}\left[3\ _{2}F_{1}\left(\frac{1}{6},\frac{1}{2},\frac{3}{2};e^{-3r}\right)-2\ _{3}F_{2}\left(\frac{1}{6},\frac{1}{2},\frac{1}{2};\frac{3}{2},\frac{3}{2};e^{-3r}\right)\right],\nonumber \\
y & = & C_{2}-\coth\left(\frac{3}{2}r\right)\left(C_{1}+C_{2}\left(1+\frac{3}{2}r\right)\right)+z+q_{b}^{2}w,\nonumber \\
z & = & -\frac{e^{-9r/2}\left(1+e^{-3r}\right)\left(9e^{3r}\ _{3}F_{2}\left(\frac{1}{2},\frac{1}{2},\frac{19}{6};\frac{3}{2},\frac{3}{2};e^{-3r}\right)+\ _{3}F_{2}\left(\frac{3}{2},\frac{3}{2},\frac{19}{6};\frac{5}{2},\frac{5}{2};e^{-3r}\right)\right)}{162\left(1-e^{-3r}\right)}\nonumber \\
 &  & -\frac{8e^{-3r/2}\left(3+10e^{-3r}\right)\ _{2}F_{1}\left(\frac{1}{6},\frac{1}{2},\frac{3}{2},e^{-3r}\right)}{819\left(1-e^{-3r}\right)}+\frac{e^{-15r/2}\left(38e^{3r}+8e^{6r}-40\right)}{273\left(1-e^{-3r}\right)^{13/6}},\nonumber \\
w & = & \frac{e^{-3r/2}\left(2e^{-3r}+1\right)}{7\left(1-e^{-3r}\right)^{7/6}}-\frac{e^{-3r/2}\left(4e^{-3r}+3\right)\ _{2}F_{1}\left(\frac{1}{6},\frac{1}{2},\frac{1}{2},e^{-3r}\right)}{21\left(1-e^{-3r}\right)}\nonumber \\
 &  & -\frac{e^{-3r/2}\left(1+e^{-3r}\right)\left(e^{-6r}\ _{3}F_{2}\left(\frac{5}{2},\frac{5}{2},\frac{19}{6};\frac{7}{2},\frac{7}{2};e^{-3r}\right)-25\ _{3}F_{2}\left(\frac{1}{2},\frac{1}{2},\frac{19}{6};\frac{3}{2},\frac{3}{2};e^{-3r}\right)\right)}{300\left(1-e^{-3r}\right)},\nonumber \\
v & = & y-6M_{2}+3\coth\left(\frac{3}{2}r\right)\left(M_{2}\left(3r+2\right)+M_{1}\right)-5z+\frac{q_{b}^{2}}{3}w.\label{eq:24}
\end{eqnarray}

\noindent Here $A_{1,2},B_{1,2},M_{1,2},C_{1,2}$ are eight integration
constants and some of them could be determined by some physical requirements.
For example, the zero-energy constraint (\ref{eq:13}) provides a
condition to the first order in $\epsilon_{f}$, which is

\begin{equation}
5B_{1}-B_{2}-18\left(C_{2}+4M_{2}\right)=0.\label{eq:25}
\end{equation}

\subsubsection*{Asymptotics}

\noindent Other constraints for the integration constants in (\ref{eq:23})
(\ref{eq:24}) would arise by analyzing the asymptotics of this solution.
Since our solution is a perturbation to the zero-th order solution
(\ref{eq:16}), it should be regularity at the tip of the ($x_{4},u$)
cigar which corresponds to the limit of $r\rightarrow\infty$ (i.e.
it gives IR behavior). As a comparison with \cite{key-18}, we work
in $x$ coordinate and obtain the following IR asymptotics ($r\rightarrow\infty$
i.e. $x=e^{-3r/2}\rightarrow0$),

\begin{eqnarray}
\lambda_{1} & = & \frac{3\left(A_{1}-A_{2}-4C_{1}\right)-2\left(B_{1}-B_{2}-6C_{2}\right)\log\left(x\right)}{12}+\mathcal{O}\left(x^{2}\right),\nonumber \\
\widetilde{\lambda}_{1} & = & \frac{-3\left(3A_{1}+A_{2}+4C_{1}\right)+2\left(3B_{1}+B_{2}+6C_{2}\right)\log\left(x\right)}{12}+\mathcal{O}\left(x^{2}\right),\nonumber \\
\phi_{1} & = & \frac{3\left(A_{1}-5A_{2}-4C_{1}\right)+2\left(-B_{1}+5B_{2}+6C_{2}\right)\log\left(x\right)}{12}+\mathcal{O}\left(x^{2}\right),\nonumber \\
\nu_{1} & = & \frac{1}{12}\left(A_{1}-5A_{2}-12M_{1}\right)-\frac{1}{18}\left(B_{1}-5B_{2}-36M_{2}\right)\log\left(x\right)+\mathcal{O}\left(x^{2}\right).\label{eq:26}
\end{eqnarray}

\noindent Accordingly, it yields the following constraints for the
integration constants,

\begin{equation}
B_{1}=6C_{2},\ B_{2}=0,\ M_{2}=\frac{C_{2}}{6}.\label{eq:27}
\end{equation}

\noindent Note that (\ref{eq:27}) satisfies (\ref{eq:25}) automatically.
And the UV behavior of functions are given as follows ($r\rightarrow0$
i.e. $x=e^{-3r/2}\rightarrow1$),

\begin{eqnarray}
\lambda_{1} & = & \frac{-C_{1}-C_{2}+k}{1-x}+\frac{101}{455\left(2\right)^{1/6}\left(1-x\right)^{1/6}}+\lambda_{1}^{UV}+\mathcal{O}\left(1-x\right)^{5/6},\nonumber \\
\widetilde{\lambda}_{1} & = & \frac{-C_{1}-C_{2}+k}{1-x}-\frac{29}{455\left(2\right)^{1/6}\left(1-x\right)^{1/6}}+\widetilde{\lambda}_{1}^{UV}+\mathcal{O}\left(1-x\right)^{5/6},\nonumber \\
\phi_{1} & = & \frac{-C_{1}-C_{2}+k}{1-x}+\frac{361}{455\left(2\right)^{1/6}\left(1-x\right)^{1/6}}+\phi_{1}^{UV}+\mathcal{O}\left(1-x\right)^{5/6},\nonumber \\
\nu_{1} & = & \frac{-M_{1}-2M_{2}+K}{1-x}+\frac{25}{91\left(2\right)^{1/6}\left(1-x\right)^{1/6}}+\nu_{1}^{UV}+\mathcal{O}\left(1-x\right)^{5/6},\label{eq:28}
\end{eqnarray}

\noindent with

\begin{eqnarray}
k & = & -\frac{\pi^{3/2}\left(3+\sqrt{3}\pi+3\log\left(\frac{27}{16}\right)\right)}{130\Gamma\left(-\frac{5}{3}\right)\Gamma\left(\frac{1}{6}\right)}-\frac{3\pi^{3/2}q_{b}^{2}\left(\sqrt{3}\pi+3\log\left(\frac{27}{16}\right)+12\right)}{560\Gamma\left(-\frac{5}{3}\right)\Gamma\left(\frac{1}{6}\right)},\nonumber \\
K & = & \frac{5}{3}k+\frac{2\pi^{3/2}q_{b}^{2}\left(\sqrt{3}\pi+12+3\log\left(\frac{27}{16}\right)\right)}{567\Gamma\left(\frac{1}{6}\right)\Gamma\left(\frac{4}{3}\right)},\nonumber \\
\lambda_{1}^{UV} & = & \frac{A_{1}-A_{2}+2\left(C_{1}+C_{2}\right)}{4}+\frac{\pi^{2/3}\left(681-85\sqrt{3}\pi+1020\log\left(2\right)-765\log\left(3\right)\right)}{1092\Gamma\left(\frac{1}{6}\right)\Gamma\left(-\frac{2}{3}\right)}\nonumber \\
 &  & -\frac{q_{b}^{2}\pi^{3/2}\left(36+17\sqrt{3}\pi+51\log\left(\frac{27}{16}\right)\right)}{672\Gamma\left(\frac{1}{6}\right)\Gamma\left(-\frac{2}{3}\right)},\nonumber \\
\widetilde{\lambda}_{1}^{UV} & = & \frac{2\left(C_{1}+C_{2}\right)-3A_{1}-A_{2}}{4}+\frac{\pi^{2/3}\left(-255+19\sqrt{3}\pi+57\log\left(\frac{27}{16}\right)\right)}{1092\Gamma\left(\frac{1}{6}\right)\Gamma\left(-\frac{2}{3}\right)}\nonumber \\
 &  & -\frac{q_{b}^{2}\pi^{3/2}\left(36+17\sqrt{3}\pi+51\log\left(\frac{27}{16}\right)\right)}{672\Gamma\left(\frac{1}{6}\right)\Gamma\left(-\frac{2}{3}\right)},\nonumber \\
\phi_{1}^{UV} & = & \frac{A_{1}-5A_{2}+2\left(C_{1}+C_{2}\right)}{4}-\frac{6559693\sqrt{3}\Gamma\left(-\frac{49}{6}\right)}{30441996288\Gamma\left(-\frac{23}{3}\right)}\big(-2553+293\sqrt{3}\pi-3516\log\left(2\right)\nonumber \\
 &  & +2637\log\left(3\right)\big)-\frac{q_{b}^{2}\sqrt{\pi}85276009\Gamma\left(-\frac{49}{6}\right)}{243535970304\Gamma\left(-\frac{23}{3}\right)}\left(36+73\sqrt{3}\pi-876\log\left(2\right)+657\log\left(3\right)\right),\nonumber \\
\nu_{1}^{UV} & = & \frac{A_{1}-5A_{2}+6M_{1}+12M_{2}}{12}-\frac{q_{b}^{2}\pi^{3/2}\left(-4+23\sqrt{3}\pi+69\log\left(\frac{27}{16}\right)\right)}{672\Gamma\left(\frac{1}{6}\right)\Gamma\left(-\frac{2}{3}\right)}\nonumber \\
 &  & +\frac{344080\pi^{3/2}\Gamma\left(-\frac{23}{3}\right)}{808321761\Gamma\left(-\frac{5}{3}\right)^{2}\Gamma\left(-\frac{2}{3}\right)^{2}\Gamma\left(\frac{1}{6}\right)^{2}}\left(-823+107\sqrt{3}\pi+420\log\left(3\right)\right)\times\nonumber \\
 &  & \left[-34875948800\pi^{3/2}\Gamma\left(-\frac{23}{3}\right)\log\left(\frac{27}{16}\right)\log\left(2\right)-28431\Gamma\left(-\frac{5}{3}\right)\Gamma\left(-\frac{2}{3}\right)\Gamma\left(\frac{1}{6}\right)\right],\nonumber \\
\label{eq:29}
\end{eqnarray}

\noindent The sub-leading terms in (\ref{eq:28}), diverging as $\left(1-x\right)^{-1/6}$,
do not depend on any integration constants, which are same as in \cite{key-18}
and could be interpreted as the dual of the ``universal'' terms.
In the UV asymptotics, the combinations of the appearing integration
constants may be interpreted as corresponding to some gauge invariant
operators, however it is less clear about what the combinations of
these functions correspond to gauge invariant operators. Nevertheless,
in order to omit the sources or VEVs of the dual operators, at least
to switch off the most divergent terms in (\ref{eq:29}), we impose
the prudent condition as \cite{key-18}, 

\begin{equation}
C_{1}+C_{2}=k,\ M_{1}+2M_{2}=K.\label{eq:30}
\end{equation}

\noindent And we do not have any more constraints on the integration
constants appearing in our solution, thus the integration constants
$A_{1},A_{2},C_{2}$ could not be determined here and we have to keep
them generic. 

In principle, the integration constants should be determined by analyzing
the complete D4-$\mathrm{D}8/\overline{\mathrm{D}8}$ solution of
this model. The full D4-$\mathrm{D}8/\overline{\mathrm{D}8}$ solution
must depend on the physical values of D4 and $\mathrm{D}8/\overline{\mathrm{D}8}$-branes.
Accordingly, if we expand the complete D4-$\mathrm{D}8/\overline{\mathrm{D}8}$
solution in the Veneziano limit, there must be some constants depending
on the physical values of $\mathrm{D}8/\overline{\mathrm{D}8}$-branes
additional to the unflavored D4-brane solution as zero-th order solution.
Therefore these extra constants should correspond to the integration
constants presented in our gravitational solution where the flavored
backreaction is perturbation. While this is the standard way to fix
the integration constants, the complete D4-$\mathrm{D}8/\overline{\mathrm{D}8}$
solution is currently out of reach. However, at least it is easy to
understand that the integration constant must depend on the magnetic
field ($q_{b}$ as another constant as the input of our theory). 

Although this is a bit different from the case without the magnetic
field in \cite{key-18}, a possibly special choice of $A_{1},A_{2},C_{2}$
may remain as \cite{key-18}, which is

\begin{equation}
A_{1}=\frac{81\sqrt{3}\pi^{2}\left(-9+\sqrt{3}\pi-12\log2+9\log3\right)}{43120\left(2\right)^{2/3}\Gamma\left(-\frac{14}{3}\right)\Gamma\left(-\frac{2}{3}\right)^{2}},\ A_{2}=-2A_{1}\ C_{2}=0.\label{eq:31}
\end{equation}
Since our solution is based on the expansion of small $q_{b}$, we
can, for example, fix $C_{2}=0,\ A_{2}=-2A_{1}$ in (\ref{eq:31})
and look for the relations between $A_{1}$ and $q_{b}$ if necessary.
These integration constants may be further determined when we study
the thermodynamics as in Section 6.

\subsection{Deconfined geometry}

The deconfined background geometry of this model in unflavored case
corresponds to the black D4-brane solution. The circle $x_{4}$ never
shrinks while the Euclideanized temporal circle shrinks at $u=u_{T}$.
The flavor branes take the position at $x_{4}=const.$ and the configuration
of a stack of parallel $\mathrm{D}8/\overline{\mathrm{D}8}$-branes
is recognized as the chirally symmetric phase in the dual field theory.

\subsubsection*{Ansatz and solution}

Similarly as the confined case, we turn on a constant $U\left(1\right)$
gauge field strength as a background magnetic field on the flavor
branes and consider two stacks of flavor branes smeared on the $x_{4}$
circle. The relevant action (with the flavor branes putting onshell)
reads as (\ref{eq:10}). We use the following ansatz for the metric
in string frame as,

\begin{equation}
ds^{2}=-e^{2\widetilde{\lambda}}dt^{2}+e^{2\lambda}dx_{a}dx^{a}+e^{2\lambda_{s}}dx_{4}^{2}+l_{s}^{2}e^{-2\varphi}d\rho^{2}+l_{s}^{2}e^{2\nu}d\Omega_{4}^{2},\label{eq:32}
\end{equation}

\noindent where

\begin{equation}
\varphi=2\phi-3\lambda-\widetilde{\lambda}-\lambda_{s}-4\nu.
\end{equation}

\noindent And we also adopt the ansatz for the gauge field strength
as $2\pi\alpha^{\prime}F_{12}=b$ as the confined case. Here $b$
also represents a dimensionless constant. Inserting the ansatz (\ref{eq:32})
and the magnetic field into (\ref{eq:10}), it yields the following
1d action,

\begin{eqnarray}
S & = & \mathcal{V}\int d\rho\left[-3\dot{\lambda}^{2}-\dot{\lambda}_{s}^{2}-\dot{\widetilde{\lambda}}^{2}-4\dot{\nu}^{2}+\dot{\varphi}^{2}+V+\mathrm{total\ derivative}\right],\nonumber \\
V & = & 12e^{-2\nu-2\varphi}-Q_{c}^{2}e^{3\lambda+\lambda_{s}+\widetilde{\lambda}-4\nu-\varphi}-Q_{f}e^{\frac{3}{2}\lambda-\frac{1}{2}\lambda_{s}+\frac{1}{2}\widetilde{\lambda}+2\nu-\frac{3}{2}\varphi}\sqrt{1+b^{2}e^{-4\lambda}}.\label{eq:34}
\end{eqnarray}

\noindent Similarly, this action (\ref{eq:34}) should also be supported
by the zero-energy constraint as (\ref{eq:13}). Then we can obtain
the equations of motion as (derivatives are w.r.t. $\rho$)

\begin{eqnarray}
\ddot{\lambda}-\frac{Q_{c}^{2}}{2}e^{6\lambda+2\lambda_{s}+2\widetilde{\lambda}-2\phi} & = & \frac{Q_{f}}{12}\left(3-b^{2}e^{-4\lambda}\right)\frac{e^{6\lambda+\lambda_{s}+2\widetilde{\lambda}-3\phi+8\nu}}{\sqrt{1+b^{2}e^{-4\lambda}}},\nonumber \\
\ddot{\lambda}_{s}-\frac{Q_{c}^{2}}{2}e^{6\lambda+2\lambda_{s}+2\widetilde{\lambda}-2\phi} & = & -\frac{Q_{f}}{4}\left(1+b^{2}e^{-4\lambda}\right)\frac{e^{6\lambda+\lambda_{s}+2\widetilde{\lambda}-3\phi+8\nu}}{\sqrt{1+b^{2}e^{-4\lambda}}},\nonumber \\
\ddot{\widetilde{\lambda}}-\frac{Q_{c}^{2}}{2}e^{6\lambda+2\lambda_{s}+2\widetilde{\lambda}-2\phi} & = & \frac{Q_{f}}{4}\left(1+b^{2}e^{-4\lambda}\right)\frac{e^{6\lambda+\lambda_{s}+2\widetilde{\lambda}-3\phi+8\nu}}{\sqrt{1+b^{2}e^{-4\lambda}}}\nonumber \\
\ddot{\nu}+\frac{Q_{c}^{2}}{2}e^{6\lambda+2\lambda_{s}+2\widetilde{\lambda}-2\phi}-3e^{6\lambda+2\lambda_{s}+2\widetilde{\lambda}-4\phi+6\nu} & = & \frac{Q_{f}}{4}\left(1+b^{2}e^{-4\lambda}\right)\frac{e^{6\lambda+\lambda_{s}+2\widetilde{\lambda}-3\phi+8\nu}}{\sqrt{1+b^{2}e^{-4\lambda}}},\nonumber \\
\ddot{\phi}-\frac{Q_{c}^{2}}{2}e^{6\lambda+2\lambda_{s}+2\widetilde{\lambda}-2\phi} & = & \frac{Q_{f}}{4}\left(5+3b^{2}\right)\frac{e^{6\lambda+\lambda_{s}+2\widetilde{\lambda}-3\phi+8\nu}}{\sqrt{1+b^{2}e^{-4\lambda}}}.\nonumber \\
\label{eq:35}
\end{eqnarray}

\noindent Since we are going to search for a perturbative solution
in the first order of $N_{f}/N_{c}$, we choose the zero-th order
solution as the unflavored solution for deconfined case, which is

\begin{eqnarray}
\lambda_{0}\left(r\right) & = & f_{0}\left(r\right)+\frac{3}{4}\log\left(\frac{u_{T}}{R}\right),\nonumber \\
\lambda_{s}\left(r\right) & = & \lambda_{0}\left(r\right),\nonumber \\
\widetilde{\lambda}_{0}\left(r\right) & = & f_{0}\left(r\right)-\frac{3}{2}r+\frac{3}{4}\log\left(\frac{u_{T}}{R}\right),\nonumber \\
\phi_{0}\left(r\right) & = & f_{0}\left(r\right)+\frac{3}{4}\log\left(\frac{u_{T}}{R}\right)+\log g_{s},\nonumber \\
\nu_{0}\left(r\right) & = & \frac{1}{3}f_{0}\left(r\right)+\frac{1}{4}\log\left(\frac{u_{T}}{R}\right)+\log\left(\frac{R}{l_{s}}\right),\label{eq:36}
\end{eqnarray}

\noindent where we have defined

\begin{eqnarray}
f_{0}\left(r\right) & = & -\frac{1}{4}\log\left[1-e^{-3r}\right],
\end{eqnarray}

\noindent and

\begin{eqnarray}
r=a_{T}\rho & , & \ a_{T}=\frac{\sqrt{2}Q_{c}u_{T}^{3}}{3R^{3}g_{s}}=\frac{u_{T}^{3}}{l_{s}^{3}g_{s}},\ q_{b}=\frac{R^{3/2}}{u_{T}^{3/2}}b.\label{eq:38}
\end{eqnarray}

\noindent Then we expand all the fields as what we have done in the
confined case,

\begin{equation}
\Psi\left(r\right)=\Psi_{0}\left(r\right)+\epsilon_{fT}\Psi_{1}\left(r\right)+\mathcal{O}\left(\epsilon_{fT}^{2}\right),
\end{equation}

\noindent with

\begin{equation}
\epsilon_{fT}=\frac{R^{3/2}u_{T}^{1/2}g_{s}}{l_{s}^{2}}Q_{f}=\epsilon_{f}\sqrt{\frac{u_{T}}{u_{0}}}=\frac{\lambda_{4}^{2}}{12\pi^{3}}\frac{2\pi T}{M_{KK}}\frac{N_{f}}{N_{c}}\ll1
\end{equation}

\noindent Here the relation between $u_{0},\ u_{T}$ and $M_{KK},\ T$
from zero-th order solution has been imposed. And we have required
that $\epsilon_{fT}=\epsilon_{f}$ at the phase transition which thus
suggests a definition of running coupling as \cite{key-18}. Then
the equations of motion for the leading order functions used in the
metric are (derivatives are w.r.t. $r$),

\begin{eqnarray}
\lambda_{1}^{\prime\prime}-\frac{9}{2}\frac{e^{-3r}}{\left(1-e^{-3r}\right)^{2}}\left(3\lambda_{1}+\lambda_{s1}+\widetilde{\lambda}_{1}-\phi_{1}\right) & = & \frac{1}{4}\frac{e^{-3r}}{\left(1-e^{-3r}\right)^{13/6}}\frac{1}{\sqrt{1+\left(1-e^{-3r}\right)q_{b}^{2}}}\nonumber \\
 &  & -\frac{q_{b}^{2}}{12}\frac{e^{-3r}}{\left(1-e^{-3r}\right)^{7/6}}\frac{1}{\sqrt{1+\left(1-e^{-3r}\right)q_{b}^{2}}},\nonumber \\
\lambda_{s1}^{\prime\prime}-\frac{9}{2}\frac{e^{-3r}}{\left(1-e^{-3r}\right)^{2}}\left(3\lambda_{1}+\lambda_{s1}+\widetilde{\lambda}_{1}-\phi_{1}\right) & = & -\frac{1}{4}\frac{e^{-3r}}{\left(1-e^{-3r}\right)^{13/6}}\frac{1}{\sqrt{1+\left(1-e^{-3r}\right)q_{b}^{2}}}\nonumber \\
 &  & -\frac{q_{b}^{2}}{4}\frac{e^{-3r}}{\left(1-e^{-3r}\right)^{7/6}}\frac{1}{\sqrt{1+\left(1-e^{-3r}\right)q_{b}^{2}}},\nonumber \\
\widetilde{\lambda}_{1}^{\prime\prime}-\frac{9}{2}\frac{e^{-3r}}{\left(1-e^{-3r}\right)^{2}}\left(3\lambda_{1}+\lambda_{s1}+\widetilde{\lambda}_{1}-\phi_{1}\right) & = & \frac{1}{4}\frac{e^{-3r}}{\left(1-e^{-3r}\right)^{13/6}}\frac{1}{\sqrt{1+\left(1-e^{-3r}\right)q_{b}^{2}}}\nonumber \\
 &  & +\frac{q_{b}^{2}}{4}\frac{e^{-3r}}{\left(1-e^{-3r}\right)^{7/6}}\frac{1}{\sqrt{1+\left(1-e^{-3r}\right)q_{b}^{2}}},\nonumber \\
\nu_{1}^{\prime\prime}-\frac{3}{2}\frac{e^{-3r}}{\left(1-e^{-3r}\right)^{2}}\left(3\lambda_{1}+\lambda_{s1}+\widetilde{\lambda}_{1}-5\phi_{1}+12\nu_{1}\right) & = & \frac{1}{4}\frac{e^{-3r}}{\left(1-e^{-3r}\right)^{13/6}}\frac{1}{\sqrt{1+\left(1-e^{-3r}\right)q_{b}^{2}}}\nonumber \\
 &  & +\frac{q_{b}^{2}}{4}\frac{e^{-3r}}{\left(1-e^{-3r}\right)^{7/6}}\frac{1}{\sqrt{1+\left(1-e^{-3r}\right)q_{b}^{2}}},\nonumber \\
\phi_{1}^{\prime\prime}-\frac{9}{2}\frac{e^{-3r}}{\left(1-e^{-3r}\right)^{2}}\left(3\lambda_{1}+\lambda_{s1}+\widetilde{\lambda}_{1}-\phi_{1}\right) & = & \frac{5}{4}\frac{e^{-3r}}{\left(1-e^{-3r}\right)^{13/6}}\frac{1}{\sqrt{1+\left(1-e^{-3r}\right)q_{b}^{2}}}\nonumber \\
 &  & +\frac{3q_{b}^{2}}{4}\frac{e^{-3r}}{\left(1-e^{-3r}\right)^{7/6}}\frac{1}{\sqrt{1+\left(1-e^{-3r}\right)q_{b}^{2}}}.\nonumber \\
\label{eq:41}
\end{eqnarray}

\noindent We will also focus on the case of small magnetic field instead
of solving (\ref{eq:41}) exactly in $q_{b}$, i.e. keeping $q_{b}^{2}$
terms by an expansion. So in a word, we need to solve the following
equations,

\begin{eqnarray}
\lambda_{1}^{\prime\prime}-\frac{9}{2}\frac{e^{-3r}}{\left(1-e^{-3r}\right)^{2}}\left(3\lambda_{1}+\lambda_{s1}+\widetilde{\lambda}_{1}-\phi_{1}\right) & = & \frac{1}{4}\frac{e^{-3r}}{\left(1-e^{-3r}\right)^{13/6}}-\frac{5q_{b}^{2}}{24}\frac{e^{-3r}}{\left(1-e^{-3r}\right)^{7/6}},\nonumber \\
\lambda_{s1}^{\prime\prime}-\frac{9}{2}\frac{e^{-3r}}{\left(1-e^{-3r}\right)^{2}}\left(3\lambda_{1}+\lambda_{s1}+\widetilde{\lambda}_{1}-\phi_{1}\right) & = & -\frac{1}{4}\frac{e^{-3r}}{\left(1-e^{-3r}\right)^{13/6}}-\frac{q_{b}^{2}}{8}\frac{e^{-3r}}{\left(1-e^{-3r}\right)^{7/6}},\nonumber \\
\widetilde{\lambda}_{1}^{\prime\prime}-\frac{9}{2}\frac{e^{-3r}}{\left(1-e^{-3r}\right)^{2}}\left(3\lambda_{1}+\lambda_{s1}+\widetilde{\lambda}_{1}-\phi_{1}\right) & = & \frac{1}{4}\frac{e^{-3r}}{\left(1-e^{-3r}\right)^{13/6}}+\frac{q_{b}^{2}}{8}\frac{e^{-3r}}{\left(1-e^{-3r}\right)^{7/6}},\nonumber \\
\nu_{1}^{\prime\prime}-\frac{3}{2}\frac{e^{-3r}}{\left(1-e^{-3r}\right)^{2}}\left(3\lambda_{1}+\lambda_{s1}+\widetilde{\lambda}_{1}-5\phi_{1}+12\nu_{1}\right) & = & \frac{1}{4}\frac{e^{-3r}}{\left(1-e^{-3r}\right)^{13/6}}+\frac{q_{b}^{2}}{8}\frac{e^{-3r}}{\left(1-e^{-3r}\right)^{7/6}},\nonumber \\
\phi_{1}^{\prime\prime}-\frac{9}{2}\frac{e^{-3r}}{\left(1-e^{-3r}\right)^{2}}\left(3\lambda_{1}+\lambda_{s1}+\widetilde{\lambda}_{1}-\phi_{1}\right) & = & \frac{5}{4}\frac{e^{-3r}}{\left(1-e^{-3r}\right)^{13/6}}+\frac{q_{b}^{2}}{8}\frac{e^{-3r}}{\left(1-e^{-3r}\right)^{7/6}}.\nonumber \\
\label{eq:42}
\end{eqnarray}

\noindent With the similar tricks used for the confined case, we thus
have the solution as,

\begin{eqnarray}
\lambda_{1} & = & \frac{1}{28}f+\frac{1}{24}q_{b}^{2}h+y-\frac{1}{4}\left(a_{2}-a_{1}-a_{3}\right)-\frac{1}{4}\left(b_{2}-b_{1}-b_{3}\right)r,\nonumber \\
\lambda_{s1} & = & \lambda_{1}-\frac{1}{21}f+\frac{q_{b}^{2}}{12}h-a_{1}-b_{1}r,\nonumber \\
\widetilde{\lambda}_{1} & = & \lambda_{1}+\frac{1}{3}q_{b}^{2}h-a_{3}-b_{3}r,\nonumber \\
\phi_{1} & = & \lambda_{1}+\frac{2}{21}f+\frac{1}{3}q_{b}^{2}h-a_{2,}-b_{2}r,\nonumber \\
\nu_{1} & = & w-y+\frac{11}{252}f+\frac{q_{b}^{2}}{8}h,\label{eq:43}
\end{eqnarray}

\noindent where the functions in (\ref{eq:43}) are given as

\begin{eqnarray}
f\left(r\right) & = & \frac{6}{\left(1-e^{-3r}\right)^{1/6}}+\sqrt{3}\tan^{-1}\left[\frac{2\left(1-e^{-3r}\right)^{1/6}-1}{\sqrt{3}}\right]+\sqrt{3}\tan^{-1}\left[\frac{2\left(1-e^{-3r}\right)^{1/6}+1}{\sqrt{3}}\right]\nonumber \\
 &  & 2\tanh^{-1}\left[\left(1-e^{-3r}\right)^{1/6}\right]-\coth^{-1}\left[\left(1-e^{-3r}\right)^{1/6}+\frac{1}{\left(1-e^{-3r}\right)^{1/6}}\right],\nonumber \\
h\left(r\right) & = & -\frac{4}{\left(1-e^{-3r}\right)^{1/6}}+\frac{2}{3}f\left(r\right),\nonumber \\
y\left(r\right) & = & c_{2}-\coth\left(\frac{3r}{2}\right)\left[c_{2}\left(\frac{3}{2}r+1\right)+c_{1}\right]+q_{b}^{2}i\left(r\right)+z\left(r\right),\nonumber \\
z\left(r\right) & = & \frac{3e^{3r}\left(1-e^{-3r}\right)^{5/6}-\frac{\sqrt{3}}{2}\left(e^{3r}+1\right)\left[\tan^{-1}\left(\frac{2\left(1-e^{-3r}\right)^{1/6}-1}{\sqrt{3}}\right)+\tan^{-1}\left(\frac{2\left(1-e^{-3r}\right)^{1/6}+1}{\sqrt{3}}\right)\right]}{546\left(e^{3r}-1\right)}\nonumber \\
 &  & +\frac{1}{2}\left(e^{3r}+1\right)\frac{2\tanh^{-1}\left[\left(1-e^{-3r}\right)^{1/6}\right]+\coth^{-1}\left[\left(1-e^{-3r}\right)^{1/6}+\frac{1}{\left(1-e^{-3r}\right)^{1/6}}\right]}{546\left(e^{3r}-1\right)},\nonumber \\
i\left(r\right) & = & -\frac{\left(7e^{-3r}+3\right)}{7\left(1-e^{-3r}\right)^{7/6}}+\frac{5\left(1+e^{-3r}\right)}{42\left(1-e^{-3r}\right)}f\left(r\right),\nonumber \\
w\left(r\right) & = & 2m_{2}-\left[m_{1}+\left(3r+2\right)m_{2}\right]\coth\left(\frac{3r}{2}\right)+\frac{1}{12}\left(a_{1}-5a_{2}+a_{3}+b_{1}r-5b_{2}r+b_{3}r\right)\nonumber \\
 &  & +\frac{5}{3}z\left(r\right)-q_{b}^{2}i\left(r\right)+y\left(r\right).\label{eq:44}
\end{eqnarray}
The integration constants are represented by $a_{1,2},\ b_{1,2},\ m_{1,2}$.
And the zero-energy condition (\ref{eq:13}) in the case of small
$q_{b}$ thus is 

\begin{equation}
-3\dot{\lambda}^{2}-\dot{\lambda}_{s}^{2}-\dot{\widetilde{\lambda}}^{2}-4\dot{\nu}^{2}+\dot{\varphi}^{2}-\frac{b^{2}}{2}Q_{f}e^{-\frac{5}{2}\lambda-\frac{1}{2}\lambda_{s}+\frac{1}{2}\widetilde{\lambda}+2\nu-\frac{3}{2}\varphi}-P=0,\label{eq:45}
\end{equation}

\noindent where

\begin{equation}
P=12e^{-2\nu-2\varphi}+Q_{f}e^{\frac{3}{2}\lambda-\frac{1}{2}\lambda_{s}+\frac{1}{2}\widetilde{\lambda}+2\nu-\frac{3}{2}\varphi}-Q_{c}^{2}e^{3\lambda+\lambda_{s}+\widetilde{\lambda}-4\nu-\varphi}.\label{eq:46}
\end{equation}

\noindent Notice that (\ref{eq:45}) would be satisfied with the leading
order solution if

\begin{equation}
-2b_{1}-2b_{2}+10b_{3}+3\left(-12c_{2}-48m_{2}-\frac{20}{7}q_{b}^{2}\right)=0.\label{eq:47}
\end{equation}

\subsubsection*{Asymptotics}

\noindent The near horizon (i.e. $x=e^{-3r/2}\rightarrow0$) behavior
of the relevant functions are given as follows,

\begin{eqnarray}
\lambda_{1} & \sim & const.+\frac{228-546b_{1}+546b_{2}-546b_{3}+3276c_{2}+962q_{b}^{2}}{3276}\log\left(x\right)+\mathcal{O}\left(x\right),\nonumber \\
\widetilde{\lambda}_{1} & \sim & const.+\frac{76-182b_{1}+182b_{2}+546b_{3}+1092c_{2}+442q_{b}^{2}}{1092}\log\left(x\right)+\mathcal{O}\left(x\right),\nonumber \\
\lambda_{s1} & \sim & const.+\frac{-28+546b_{1}+182b_{2}-182b_{3}+1092c_{2}+442q_{b}^{2}}{1092}\log\left(x\right)+\mathcal{O}\left(x\right),\nonumber \\
\phi_{1} & \sim & const.+\frac{284-182b_{1}+910b_{2}-182b_{3}+1092c_{2}+806q_{b}^{2}}{1092}\log\left(x\right)+\mathcal{O}\left(x\right),\nonumber \\
\nu_{1} & \sim & const.+\frac{92+2184m_{2}+182q_{b}^{2}}{1092}\log\left(x\right)+\mathcal{O}\left(x\right).
\end{eqnarray}

\noindent Furthermore, we require that the solution is regular at
the tip of the Euclidean cigar, it thus leads the following constraints

\begin{equation}
b_{1}=\frac{1}{7}-\frac{1}{6}q_{b}^{2},\ b_{2}=-\frac{2}{7}-\frac{2}{3}q_{b}^{2},\ m_{2}=\frac{5}{3276}+\frac{b_{3}}{36}+\frac{q_{b}^{2}}{216},\ c_{2}=\frac{1}{546}+\frac{b_{3}}{6}-\frac{53q_{b}^{2}}{252}.\label{eq:49}
\end{equation}

\noindent Notice that (\ref{eq:49}) fulfills the zero-energy constraint
(\ref{eq:47}) automatically as well. And the UV behavior (i.e. $x\rightarrow1$)
of these functions is,

\begin{eqnarray}
\lambda_{1} & = & -\frac{c_{1}+c_{2}}{1-x}+\frac{101}{455\left(2\right)^{1/6}\left(1-x\right)^{1/6}}+\frac{1}{4}\left[a_{1}-a_{2}+a_{3}+2\left(c_{1}+c_{2}\right)\right]+\mathcal{O}\left(\left(1-x\right)^{1/6}\right),\nonumber \\
\lambda_{s1} & = & -\frac{c_{1}+c_{2}}{1-x}+\frac{29}{455\left(2\right)^{1/6}\left(1-x\right)^{1/6}}+\frac{1}{4}\left[-3a_{1}-a_{2}+a_{3}+2\left(c_{1}+c_{2}\right)\right]+\mathcal{O}\left(\left(1-x\right)^{1/6}\right),\nonumber \\
\widetilde{\lambda}_{1} & = & -\frac{c_{1}+c_{2}}{1-x}+\frac{101}{455\left(2\right)^{1/6}\left(1-x\right)^{1/6}}+\frac{1}{4}\left[a_{1}-a_{2}-3a_{3}+2\left(c_{1}+c_{2}\right)\right]+\mathcal{O}\left(\left(1-x\right)^{1/6}\right),\nonumber \\
\phi_{1} & = & -\frac{c_{1}+c_{2}}{1-x}+\frac{361}{455\left(2\right)^{1/6}\left(1-x\right)^{1/6}}+\frac{1}{4}\left[a_{1}-5a_{2}+a_{3}+2\left(c_{1}+c_{2}\right)\right]+\mathcal{O}\left(\left(1-x\right)^{1/6}\right),\nonumber \\
\nu_{1} & = & -\frac{m_{1}+2m_{2}}{1-x}+\frac{25}{91\left(2\right)^{1/6}\left(1-x\right)^{1/6}}+\frac{1}{12}\left[a_{1}-5a_{2}+a_{3}+6m_{1}+12m_{2}\right]+\mathcal{O}\left(\left(1-x\right)^{1/6}\right).\nonumber \\
\end{eqnarray}

\noindent To eliminate the leading divergences as discussed in the
confined case, we impose

\begin{equation}
c_{1}=-c_{2},\ m_{1}=-2m_{2}.\label{eq:51}
\end{equation}
Then we do not have any more constraints for other integration constants,
thus we have to keep $a_{1,2,3}$ and $b_{3}$ generic. Nevertheless
a possible choice for $a_{1,2,3}$ and $b_{1}$ with small magnetic
field might be (same as \cite{key-18}),

\begin{equation}
a_{1}=a_{2}=a_{3}=b_{3}=0.\label{eq:52}
\end{equation}
However, we have to keep in mind that (\ref{eq:52}) is also not strictly
necessary and further determination of the integration constants will
be discussed in Section 6.

\section{Some physical properties}

In this section, we will study some holographically physical effects
in hadronic physics by using our magnetic-dependent backreaction solution
in confined case (\ref{eq:23}) (\ref{eq:24}). 

To begin with, since the ($x_{4},r$) cigar has to close smoothly
at the tip ($r\rightarrow\infty$), the relation between the parameter
$u_{0}$ and $M_{KK}$ is modified by the backreaction from the flavor
and magnetic field. Therefore we have,

\begin{equation}
M_{KK}=\frac{3}{2}\frac{u_{0}^{1/2}}{R^{3/2}}\left[1-\frac{\epsilon_{f}}{6}\left(5A_{1}-A_{2}-4C_{2}-12k-24K\right)\right].\label{eq:53}
\end{equation}
If using the special choice (\ref{eq:31}) as \cite{key-18}, we obtain
$5A_{1}-A_{2}-4C_{2}-12k-24K\approx2+\frac{2}{5}q_{b}^{2}$. Obviously,
with this choice, the length of the $x_{4}$ circle becomes larger
as the magnetic field increases. For the reader convenience, we also
give the relation between the parameter $u_{T}$, $R$ and Hawking
temperature $T$ in the deconfined case,

\begin{equation}
\frac{u_{T}}{R^{3}}=\frac{4}{9}\left(2\pi T\right)^{2}\left[1+\frac{2}{9}\epsilon_{fT}\left(1-\frac{3}{2}\left(a_{1}+a_{2}-5a_{3}\right)-b_{3}+\left(\frac{71}{42}+\frac{15}{14}\sqrt{3}\pi-\frac{15}{14}\log432\right)q_{b}^{2}\right)\right],\label{eq:54}
\end{equation}
as the metric has to be regular at the horizon of the Euclideanized
black hole as well.

Notice that we have to keep in mind all the discussions in this section
would not be strictly rigorous once the special choice (\ref{eq:31})
for the undetermined integration constants is imposed. Since all our
results should definitely return to \cite{key-18} if turning off
the magnetic field, we assume (\ref{eq:31}) (from \cite{key-18},
i.e. the non-magnetic case) is a simple choice for the undetermined
integration constants. Absolutely this is not necessary or strict
in our magnetic case. However because of the lack of the geometric
constraints for our gravitational solution and the less clear relation
between the integration constants and the magnetic field, some integration
constants are not determined in fact. So we can not conclude or compare
anything with \cite{key-18} if keeping all the undetermined constants
generic. Accordingly, we therefore impose the special choice (\ref{eq:31})
throughout the calculations in the following subsections. Consequently
our results in this section might not be strictly conclusive but they
are good comparisons with \cite{key-18}\footnote{Since our gravitational solution is magnetic-dependent, it is also
a parallel calculation to \cite{key-18} as a check.}.

\subsection{The running coupling}

In the Witten-Sakai-Sugimoto model, the Yang-Mills coupling constant
is related to the compactified circle $x_{4}$ \cite{key-34}. By
examining a D4-brane as the probe wrapped on the $x_{4}$ circle,
we obtain the running gauge coupling \cite{key-39} (the formulas
are expressed in the coordinate of $x=e^{-3r/2}$. )

\begin{equation}
\frac{1}{g_{YM,x}^{2}}=\frac{1}{2\pi l_{s}M_{KK,0}}e^{-\phi+\widetilde{\lambda}}=\frac{x}{g_{YM}^{2}}\left[1-\epsilon_{f}\left(\phi_{1}-\widetilde{\lambda}_{1}\right)\right].\label{eq:55}
\end{equation}
According to the UV behavior ($x\rightarrow1$) (\ref{eq:28}) of
the functions, we thus obtain the formula of the running coupling
which remains as \cite{key-18}, 

\begin{equation}
\frac{1}{g_{YM,x}^{2}}\simeq\frac{1}{g_{YM}^{2}}\left[1-\frac{3}{7}\epsilon_{f}\frac{2^{5/6}}{\left(1-x\right)^{1/6}}\right].\label{eq:56}
\end{equation}
Obviously, this formula is independent on the presence of the magnetic
field which seems different from QFT/QCD approach as \cite{key-40}
but in agreement with \cite{key-18}. Technically, (\ref{eq:56})
corresponds to the condition (\ref{eq:30}) we have chosen. In (\ref{eq:28})
we have omitted the most divergent terms by imposing (\ref{eq:30})
to turn off the sources or VEVs of some gauge invariant operators
in the dual field theory although some details about the holographic
correspondence here are also less clear. So the surviving divergences
in (\ref{eq:28}) are all independent on the integration constants,
which thus yields a integration-constant-independent divergence in
(\ref{eq:56}) by (\ref{eq:55}). In this sense our (\ref{eq:56})
is same as \cite{key-18} since we have chosen the same boundary conditions
for the gravitational solution while the gravitational solution itself
is actually different.

On the other hand (\ref{eq:56}) signals a Landau pole since the coupling
constant tends to diverge in the UV limit (i.e. $x\rightarrow1$)
which strongly differs from QCD in fact. There might be a simple interpretation
about the appearance of the Landau pole. As it is known the background
of this model is Witten's geometry \cite{key-34} at the limit $N_{c}\rightarrow\infty$.
In our backreaction case, we could require $N_{c}$ is large but not
infinity and $N_{f}/N_{c}\ll1$ fixed. Accordingly, the background
geometry is actually 11d ($AdS_{7}\times S^{4}$) while the 11th direction
is compacted on a cycle with a very small size (as some energy scales
in the dual field theory). Therefore the dual field theory could be
conformal upon this energy scale \cite{key-34}. So it is possible
to generate a Landau pole by adding flavors to a CFT.

Besides (\ref{eq:56}) only shows the the UV behavior ($x\rightarrow1$)
of the running coupling, but basically we can obtain the complete
relation between the running coupling and the magnetic field by using
(\ref{eq:55}). The behavior of $g_{YM,x}$ with $B$ is actually
quite ambiguous because of the presence of the undetermined integration
constants $A_{1},A_{2},C_{2}$. Due to the different behaviors in
UV limit, we can impose the special choice (\ref{eq:31}) to (\ref{eq:55}),
as a result it yields to a different behavior of $g_{YM,x}$ with
$B$ from the QFT result in \cite{key-40}. However, we need to emphasize
that this comparison with QCD is strictly significant only if the
theories with same number of colors and flavors are considered, otherwise
theories with different numbers of colors or flavors could have different
behaviors.

\subsection{QCD String tension}

The QCD string tension could be obtained by evaluating a string action.
It has turned out that, by using (\ref{eq:53}) the string tension
is given as\footnote{There also is other studies on flavor corrections to the static potential
in this model such as \cite{key-41}.},

\begin{equation}
T_{s}=\frac{1}{2\pi\alpha^{\prime}}e^{2\lambda}\bigg|_{x=0}=\frac{2}{27\pi}\lambda_{4}M_{KK}^{2}\left[1+\epsilon_{f}\left(3A_{1}-A_{2}-12K-8k\right)\right].\label{eq:57}
\end{equation}
Imposing the special choice (\ref{eq:31}), we have $3A_{1}-A_{2}-12K-8k\simeq1.13+0.28q_{b}^{2}$.
In this sense, we can naively conclude that the string tension increase
by the effect of the dynamical flavors and the presence of the magnetic
field. But our result (\ref{eq:57}) seems unrealistic if $T_{s}$
could be holographically interpreted as some QCD tensions, because
intuitively speaking the theory should confine less when more flavors
(or magnetic field) are added. However this behavior of the theory
should depend on which scheme is chosen and where some observable
is kept fixed, since theories with different $N_{f}$ are actually
different as mentioned \cite{key-18}. Nevertheless we are not clear
about whether the opposite behavior in (\ref{eq:57}) corresponds
to large $N_{c}$ limit or the choice (\ref{eq:31}) for the undetermined
integration constants in our theory. We believe a future study about
this is also needed.

\subsection{Baryon mass}

In AdS/CFT, a baryon is a wrapped D-brane on the extra dimensions
\cite{key-42,key-43}. Accordingly, a baryon vertex is a wrapped $\mathrm{D}4^{\prime}$-brane\footnote{In order to distinguish with the D4-branes which produces the back
ground geometry in this holographic system, we have used ``$\mathrm{D}4^{\prime}$-brane''
to denote a baryon vertex throughout this manuscript.} on $S^{4}$ in the Witten-Sakai-Sugimoto model. And it corresponds
to the deep IR of the dual field theory since it is localized at the
radial position i.e. the holographic direction. So with the Euclidean
version of the backreaction solution in the confined case, we can
easily read the wrapped $\mathrm{D}4^{\prime}$-brane action,

\begin{equation}
S_{D4'}^{E}=T_{4}\int dx^{0}d\Omega_{4}e^{-\phi}\sqrt{\det g_{5}}=T_{4}V_{S^{4}}l_{s}^{4}\int dx^{0}e^{\lambda+4\nu-\phi}\bigg|_{x=0}=m_{B}\int dx^{0},
\end{equation}
here $T_{4}=\left(2\pi\right)^{-4}l_{s}^{-5}$ is the tension of the
$\mathrm{D}4^{\prime}$-brane. Using our solution in the confined
case at $x=e^{-3r/2}\rightarrow0$ (i.e. the IR value of the radial
direction), we have the baryon mass which is given as

\begin{equation}
m_{B}=\frac{1}{27\pi}\lambda_{4}N_{c}M_{KK}\left[1+\epsilon_{f}\left(2A_{1}-A_{2}-8K+N\right)\right],
\end{equation}
where

\begin{equation}
N=\frac{16\pi^{3/2}\left(\sqrt{3}\pi+3+3\log\left(\frac{27}{16}\right)\right)}{195\Gamma\left(-\frac{5}{3}\right)\Gamma\left(\frac{1}{6}\right)}+\frac{2\pi^{3/2}q_{b}^{2}\left(\sqrt{3}\pi+12+3\log\left(\frac{27}{16}\right)\right)}{105\Gamma\left(\frac{1}{6}\right)\Gamma\left(\frac{4}{3}\right)}.
\end{equation}
For the special choice (\ref{eq:31}) it gives $2A_{1}-A_{2}-8K+N\simeq0.95+0.11q_{b}^{2}$.
Therefore, the baryon mass also increases by the modification of the
flavor dynamics and the presence of the magnetic field. The comments
are similar as in the previous subsections.

\section{Holographic renormalization with the magnetic field}

In this section, we are going to discuss the main subject of this
manuscript, i.e. study the thermodynamics and holographic renormalization
of this model, by our magnetic-dependent solution. 

Through the holographic formula $F=TS_{E}^{onshell}$, the Euclidean
gravity action is related to the free energy of this model. As we
are going to discuss the thermodynamics of this model, we need to
evaluate the Euclidean onshell action, taking into account the backreaction
by our magnetic-dependent solutions. And the Euclidean version of
the Type II A supergravity action could be obtained by a Wick rotation
from (\ref{eq:10}), which is 

\begin{align}
S_{E} & =-\frac{1}{2k_{0}^{2}}\int d^{10}x\sqrt{-g}\left[e^{-2\phi}\left(\mathcal{R}+4\left(\partial\phi\right)^{2}\right)-\frac{1}{2}\left|F_{4}\right|^{2}\right]\nonumber \\
 & \ \ -\frac{N_{f}T_{8}M_{KK}}{\pi}\int d^{10}x\frac{\sqrt{-\det\left(g+2\pi\alpha^{\prime}F\right)}}{\sqrt{g_{44}}}e^{-\phi}.\label{eq:61}
\end{align}
However, the onshell action (\ref{eq:61}) is divergent if inserting
our solutions in confined or deconfined case. Since we would like
to compare the free energy of this model with different backreaction
solutions, we have to renormalize the theory holographically. The
renormalized gravity action could be written as

\begin{equation}
S_{E}^{ren}=S_{E}+S_{GH}+S_{c.t}^{bulk}+S_{c.t}^{D8}.\label{eq:62}
\end{equation}
$S_{E}$ is the Euclidean version of the Type II A supergravity action
(\ref{eq:61}) and $S_{GH}$, $S_{c.t}^{bulk}$, $S_{c.t}^{D8}$ is
Gibbons-Hawking (GH) term, the bulk counterterm and the D8-brane counterterm
respectively. In string frame, they are given as\footnote{The bulk counter terms are given in \cite{key-44} and it has turned
out the bulk counterterm is not enough to cancel all the divergent
terms if the backreaction from flavor brane is considered. The counterterm
of the flavor branes in the presence of an external magnetic field
in the Sakai-Sugimoto model has been given in \cite{key-29} and it
is written as a covariant form in \cite{key-18}. Therefore we have
employed the covariant form for the smeared D8-brane counterterm in
(\ref{eq:63}).}

\begin{align}
S_{GH} & =\ -\frac{1}{k_{0}^{2}}\int d^{9}x\sqrt{h}e^{-2\phi}K,\nonumber \\
S_{c.t}^{bulk} & =\ \frac{1}{k_{0}^{2}}\left(\frac{g_{s}^{1/3}}{R}\right)\int d^{9}x\sqrt{h}\frac{5}{2}e^{-7\phi/3},\nonumber \\
S_{c.t}^{D8} & =\ \frac{Q_{f}}{k_{0}^{2}l_{s}^{2}}\int d^{9}x\frac{\sqrt{h}}{\sqrt{h_{44}}}\left[\chi_{1}\frac{R}{g_{s}^{1/3}}e^{-2\phi/3}-2\chi_{2}\frac{R^{2}}{g_{s}^{2/3}}e^{-\phi/3}\left(K-\frac{8}{3}n\cdot\nabla\phi-n\cdot\frac{\nabla\left(\sqrt{g_{44}}\right)}{\sqrt{g_{44}}}\right)\right],\label{eq:63}
\end{align}
where $\chi_{1,2}$ are two constants for the case of smeared D8-branes
and $h$ is the determinant of the metric at the UV boundary i.e.
the slice of the 10d metric fixed at $r=\varepsilon$ with $\varepsilon\rightarrow0$.
$K$ is the trace of the boundary extrinsic curvature whose explicit
form in our notation is

\begin{equation}
K=h^{MN}\nabla_{M}n_{N}=-\frac{1}{\sqrt{g}}\partial_{r}\left(\frac{\sqrt{g}}{\sqrt{g_{rr}}}\right)\bigg|_{r=\varepsilon},
\end{equation}
and

\begin{equation}
n^{M}=-\frac{\delta^{Mr}}{\sqrt{g_{rr}}}.
\end{equation}
Then we are going to evaluate all the terms in (\ref{eq:61}) and
(\ref{eq:63}) by our magnetic-dependent solutions both in confined
and deconfined case.

\subsection{Confined case}

Evaluating the action (\ref{eq:61}) and (\ref{eq:63}) by our magnetic-dependent
solution for confined case, we have the following onshell actions
(up to the first order on $\epsilon_{f}$)

\begin{align}
S_{E} & =-a\mathcal{V}\left[\frac{9}{4}-\frac{3}{2\varepsilon}+\epsilon_{f}h_{1}\right],\nonumber \\
S_{GH} & =-a\mathcal{V}\left[-\frac{7}{4}+\frac{19}{6\varepsilon}+\epsilon_{f}h_{2}\right],\nonumber \\
S_{c.t.}^{bulk} & =a\mathcal{V}\left[\frac{5}{3\varepsilon}+\epsilon_{f}h_{3}\right],\nonumber \\
S_{c.t.}^{D8} & =a\mathcal{V}\epsilon_{f}\left[\frac{2\chi_{1}^{c}-8\chi_{2}^{c}}{3\left(3\right)^{1/6}\varepsilon^{7/6}}+\frac{14\chi_{1}^{c}-8\chi_{2}^{c}}{12\left(3\right)^{1/6}\varepsilon^{1/6}}\right].\label{eq:66}
\end{align}
where

\begin{align}
a\mathcal{V} & =\frac{1}{2k_{0}^{2}g_{s}^{2}}\frac{V_{3}}{T}\frac{2\pi}{M_{KK}}V_{S^{4}}u_{0}^{3},\nonumber \\
h_{1} & =9C_{2}-\frac{150\pi^{3/2}}{7\Gamma\left(-\frac{2}{3}\right)\Gamma\left(\frac{1}{6}\right)}+\frac{8645\pi^{3/2}q_{b}^{2}}{2592\Gamma\left(\frac{4}{3}\right)\Gamma\left(\frac{25}{6}\right)}-\frac{823}{1365\left(3\right)^{1/6}\varepsilon^{7/6}}-\frac{6685+3330q_{b}^{2}}{924\left(3\right)^{1/6}\varepsilon^{1/6}},\nonumber \\
h_{2} & =-7C_{2}-\frac{66\pi^{3/2}}{5\Gamma\left(-\frac{5}{3}\right)\Gamma\left(\frac{1}{6}\right)}-\frac{21\pi^{3/2}q_{b}^{2}}{4\Gamma\left(-\frac{5}{3}\right)\Gamma\left(\frac{1}{6}\right)}+\frac{25}{39\left(3\right)^{1/6}\varepsilon^{7/6}}+\frac{6713+2490q_{b}^{2}}{924\left(3\right)^{1/6}\varepsilon^{1/6}},\nonumber \\
h_{3} & =\frac{14}{117\left(3\right)^{1/6}\varepsilon^{7/6}}+\frac{245+150q_{b}^{2}}{1386\left(3\right)^{1/6}\varepsilon^{1/6}}.\label{eq:67}
\end{align}
In order to cancel all the divergences in (\ref{eq:66}) and (\ref{eq:67}),
we have to choose 

\begin{equation}
\chi_{1}^{c}=-\frac{1893+15275q_{b}^{2}}{15015},\ \chi_{2}^{c}=-\frac{56+15275q_{b}^{2}}{60060}.\label{eq:68}
\end{equation}
As we can see from (\ref{eq:67}), there are magnetic-dependent divergences.
However, for a constant magnetic field, it is possible to choose the
$q_{b}$-dependent constants (\ref{eq:68}) in the counterterm action
\cite{key-29}. Therefore the renormalized action for the backreaction
case reads

\begin{equation}
S_{E}^{ren}=-\frac{1}{2}a\mathcal{V}\left(1+\epsilon_{f}\left[4C_{2}+\frac{8\pi^{3/2}}{7\Gamma\left(-\frac{2}{3}\right)\Gamma\left(\frac{1}{6}\right)}+\frac{8645\pi^{3/2}q_{b}^{2}}{5832\Gamma\left(\frac{4}{3}\right)\Gamma\left(\frac{25}{6}\right)}\right]\right).\label{eq:69}
\end{equation}

\subsection{Deconfined case}

As in the confined case, the Euclidean version of the onshell action
(plus the GH term) (which is the Gibbs free energy) is also divergent,
thus it must be renormalized by approaching the counterterms in (\ref{eq:62}).
The functional form of each term in (\ref{eq:62}) takes the same
formulas as (\ref{eq:61}) and (\ref{eq:63}) respectively, however
it needs to be evaluated by our deconfined solution. Therefore we
have,

\begin{align}
S_{E} & =-a_{T}\mathcal{V}\left[\frac{9}{4}-\frac{3}{2\varepsilon}+\epsilon_{fT}g_{1}\right],\nonumber \\
S_{GH} & =-a_{T}\mathcal{V}\left[-\frac{7}{4}+\frac{19}{6\varepsilon}+\epsilon_{fT}g_{2}\right],\nonumber \\
S_{c.t.}^{bulk} & =a_{T}\mathcal{V}\left[\frac{5}{3\varepsilon}+\epsilon_{fT}g_{3}\right],\nonumber \\
S_{c.t.}^{D8} & =a_{T}\mathcal{V}\epsilon_{fT}\left[\frac{2\chi_{1}^{d}-8\chi_{1}^{d}}{3\left(3\right)^{1/6}\varepsilon^{7/6}}+\frac{\chi_{1}^{d}-16\chi_{1}^{d}}{6\left(3\right)^{1/6}\varepsilon^{1/6}}\right].\label{eq:70}
\end{align}
where

\begin{align}
a_{T}\mathcal{V} & =\frac{1}{2k_{0}^{2}g_{s}^{2}}\frac{V_{3}}{T}\frac{2\pi}{M_{KK}}V_{S^{4}}u_{T}^{3},\nonumber \\
g_{1} & =\frac{25}{14}+\frac{3}{2}b_{3}+\left(\frac{153}{28}-\frac{5\sqrt{3}\pi}{14}+\frac{5}{7}\log12+\frac{5}{14}\log3\right)q_{b}^{2}-\frac{823}{1365\left(3\right)^{1/6}\varepsilon^{7/6}}\nonumber \\
 & \ \ -\frac{955+5562q_{b}^{2}}{924\left(3\right)^{1/6}\varepsilon^{1/6}},\nonumber \\
g_{2} & =-\frac{11}{6}-\frac{7b_{3}}{6}-\frac{133}{36}q_{b}^{2}+\frac{25}{39\left(3\right)^{1/6}\varepsilon^{7/6}}+\frac{959+3858q_{b}^{2}}{924\left(3\right)^{1/6}\varepsilon^{1/6}},\nonumber \\
g_{3} & =\frac{14}{117\left(3\right)^{1/6}\varepsilon^{7/6}}+\frac{35-1146q_{b}^{2}}{1386\left(3\right)^{1/6}\varepsilon^{1/6}}.\label{eq:71}
\end{align}
As we can see, the ``bulk counterterm'' $S_{c.t.}^{bulk}$ cancels
the $\mathcal{O}\left(\epsilon_{fT}^{0}\right)$ divergences only
as the confined case. We thus have introduced the additional counterterm,
i.e. the ``flavor counterterm'' $S_{c.t.}^{D8}$ which is related
to the D8-branes, to cancel the remaining divergences\footnote{As another possibility, to cancel the divergences is to subtract the
onshell value of $S_{E}+S_{GH}$, the value of the same combination
on some background as being a reference. }. Consequently, we have to choose the following values,

\begin{equation}
\chi_{1}^{d}=-\frac{1821-30550q_{b}^{2}}{15015},\ \chi_{2}^{d}=\frac{8+15275q_{b}^{2}}{30030},
\end{equation}
to cancel all the divergences in (\ref{eq:70}) and (\ref{eq:71}).
With these choices, we have the renormalized action in the deconfined
case which is,

\begin{equation}
S_{E}^{ren}=-\frac{1}{2}a_{T}\mathcal{V}\left[1+\epsilon_{fT}\left(-\frac{2}{21}+\frac{2}{3}b_{3}+\left(\frac{15}{7}\log3+\frac{20}{7}\log2-\frac{5\sqrt{3}\pi}{7}+\frac{223}{63}\right)q_{b}^{2}\right)\right].\label{eq:73}
\end{equation}

\section{The phase diagram}

In this section, let us discuss the phase diagram of this holographic
model in the presence of a magnetic field and compare the diagram
with lattice QCD. Since there is no chemical potential through our
setup, we will thus focus on the case of finite temperature and zero
chemical potential in QCD.

\subsection{The probe approximation}

Since our goal is to quantify the effects from the flavors on the
critical temperature in the presence of the magnetic field when the
phase transition happens between the confined and deconfined geometric
phase. Therefore, we just need to compare the free energy from the
renormalized onshell action. And we should first calculate the pressure
$p$ both for confined and deconfined phase by using,

\begin{equation}
p=-\frac{S_{E}^{ren}T}{V_{3}}.
\end{equation}
Since we have introduced the additional boundary terms for the flavor
branes in (\ref{eq:63}), it admits the holographically renormalized
bulk action. Moreover, according to our calculations, it is obvious
that the backreaction from the flavors and magnetic field is a perturbation
to the bulk geometry. So that going back to the case of the probe
limit (which means the flavor branes are treated as probes as usually
discussed in this model) should be required definitely from our gravity
solutions. Hence in the probe approximation, we have the onshell D8-brane
action with the U-shape embedding i.e. $x_{4}=const.$, which is

\begin{equation}
S_{D8}^{\mathrm{conf.}}=\frac{Q_{f}}{2k_{0}^{2}}\frac{2\pi}{l_{s}^{2}M_{KK}}\int d^{9}xe^{-\phi_{0}}\sqrt{\det g_{9}^{(0)}}=a\mathcal{V}\epsilon_{f}d_{probe}^{\mathrm{conf}},\label{eq:75}
\end{equation}
where,

\begin{equation}
d_{probe}^{\mathrm{conf.}}=\frac{2}{21\left(3\right)^{1/6}\varepsilon^{7/6}}+\frac{7+6q_{b}^{2}}{6\left(3\right)^{1/6}\varepsilon^{1/6}}-\frac{2\pi^{3/2}\left(8+7q_{b}^{2}\right)}{21\Gamma\left(\frac{1}{6}\right)\Gamma\left(\frac{4}{3}\right)}.\label{eq:76}
\end{equation}
Here we have expanded the action in small $q_{b}$ limit. It can be
found there are two divergent terms in the onshell action (\ref{eq:75})
in the UV. Therefore, in order to cancel the divergences in (\ref{eq:75}),
we have to choose

\begin{equation}
\chi_{1}^{c}=-\frac{8}{7}-q_{b}^{2},\ \chi_{2}^{c}=-\frac{1}{4}-\frac{1}{4}q_{b}^{2},
\end{equation}
in the D8-brane counterterm (\ref{eq:66}) for the probe approximation.
In a word, we obtain the renormalized action (bulk plus flavor brane)
in the probe approximation as,

\begin{equation}
S_{E,\ probe}^{ren,\mathrm{conf.}}=-\frac{1}{2}a\mathcal{V}\left[1+\epsilon_{f}\left(\frac{32\pi^{3/2}}{21\Gamma\left(\frac{1}{6}\right)\Gamma\left(\frac{4}{3}\right)}+\frac{4\pi^{3/2}}{3\Gamma\left(\frac{1}{6}\right)\Gamma\left(\frac{4}{3}\right)}q_{b}^{2}\right)\right].
\end{equation}
Accordingly, we have the pressure in the probe approximation for the
confined case,

\begin{equation}
p_{\mathrm{conf}.}^{probe}=\frac{2N_{c}^{2}\lambda_{4}M_{KK}^{4}}{3^{7}\pi^{2}}\left[1+\epsilon_{f}\left(\frac{32\pi^{3/2}}{21\Gamma\left(\frac{1}{6}\right)\Gamma\left(\frac{4}{3}\right)}+\frac{4\pi^{3/2}}{3\Gamma\left(\frac{1}{6}\right)\Gamma\left(\frac{4}{3}\right)}q_{b}^{2}\right)\right].\label{eq:79}
\end{equation}
For the deconfined case (similarly as in the confined phase), we also
have the onshell D8-brane action with the parallel embedding, which
is

\begin{equation}
S_{D8}^{\mathrm{deconf}}=\frac{Q_{f}}{2k_{0}^{2}}\frac{2\pi}{l_{s}^{2}M_{KK}}\int d^{9}xe^{-\phi_{0}}\sqrt{\det g_{9}^{(0)}}=a_{T}\mathcal{V}\epsilon_{fT}d_{probe}^{deconf},
\end{equation}
where

\begin{equation}
d_{probe}^{\mathrm{deconf}}=\frac{2}{21\left(3\right)^{1/6}\varepsilon^{7/6}}+\frac{1+6q_{b}^{2}}{6\left(3\right)^{1/6}\varepsilon^{1/6}}-\frac{2}{7}-q_{b}^{2}.
\end{equation}
So we need the following choice,

\begin{equation}
\chi_{1}^{d}=\frac{1}{7}-2q_{b}^{2},\ \chi_{2}^{d}=\frac{1}{14}-\frac{1}{2}q_{b}^{2},
\end{equation}
for the additional flavor brane counterterm in (\ref{eq:70}). Obviously,
in the probe approximation the renormalized onshell D8-brane action
reads

\begin{equation}
S_{E,\ probe}^{ren,\mathrm{deconf}}=-\frac{1}{2}a_{T}\mathcal{V}\left[1+\epsilon_{fT}\left(\frac{4}{7}+2q_{b}^{2}\right)\right].
\end{equation}
And its pressure is,

\begin{equation}
p_{\mathrm{deconf}.}^{probe}=\frac{128\pi^{4}N_{c}^{2}T^{6}\lambda_{4}}{2187M_{KK}^{2}}\left[1+\epsilon_{fT}\left(\frac{4}{7}+2q_{b}^{2}\right)\right].\label{eq:84}
\end{equation}
Consequently, we can obtain the phase diagram in the probe approximation
by comparing the pressure (\ref{eq:79}) and (\ref{eq:85}) with the
equation $p_{\mathrm{deconf}.}(T=T_{c})=p_{\mathrm{conf}.}$\footnote{At the phase transition, we have set $\epsilon_{f}=\epsilon_{fT}$
since the contribution form $\mathcal{O}\left(N_{f}/N_{c}\right)$
in $\epsilon_{fT}$ could be neglected.}, it gives

\begin{equation}
\frac{2\pi T_{c}}{M_{KK}}=1-\frac{1}{126\pi^{3}}\lambda_{4}^{2}\frac{N_{f}}{N_{c}}\left[1-\frac{8\pi^{3/2}}{3\Gamma\left(\frac{1}{6}\right)\Gamma\left(\frac{4}{3}\right)}\right]-\frac{81}{16\pi}\frac{N_{f}}{N_{c}}\frac{B^{2}}{M_{KK}^{4}}\left[1-\frac{2\pi^{3/2}}{3\Gamma\left(\frac{1}{6}\right)\Gamma\left(\frac{4}{3}\right)}\right],\label{eq:85}
\end{equation}
where

\begin{align}
1-\frac{8\pi^{3/2}}{3\Gamma\left(\frac{1}{6}\right)\Gamma\left(\frac{4}{3}\right)} & \simeq-1.987,\ \ 1-\frac{2\pi^{3/2}}{3\Gamma\left(\frac{1}{6}\right)\Gamma\left(\frac{4}{3}\right)}\simeq0.253.\label{eq:86}
\end{align}
So from (\ref{eq:85}) and (\ref{eq:86}), we could conclude that,
at zero chemical potential without the magnetic field, the critical
temperature increases by the effect of the flavors\footnote{Without the magnetic field, (\ref{eq:86}) is quantitative same as
\cite{key-18} definitely.} (see also \cite{key-18,key-29}). And we also notice that the contribution
from the magnetic field is quadratic for any $N_{f}$. Moreover, (\ref{eq:85})
shows $T_{c}$ decreases when $B$ increases (as shown in Figure 2)
which is in agreement with the lattice QCD results \cite{key-1,key-2}. 

\begin{figure}
\begin{centering}
\includegraphics[scale=0.15]{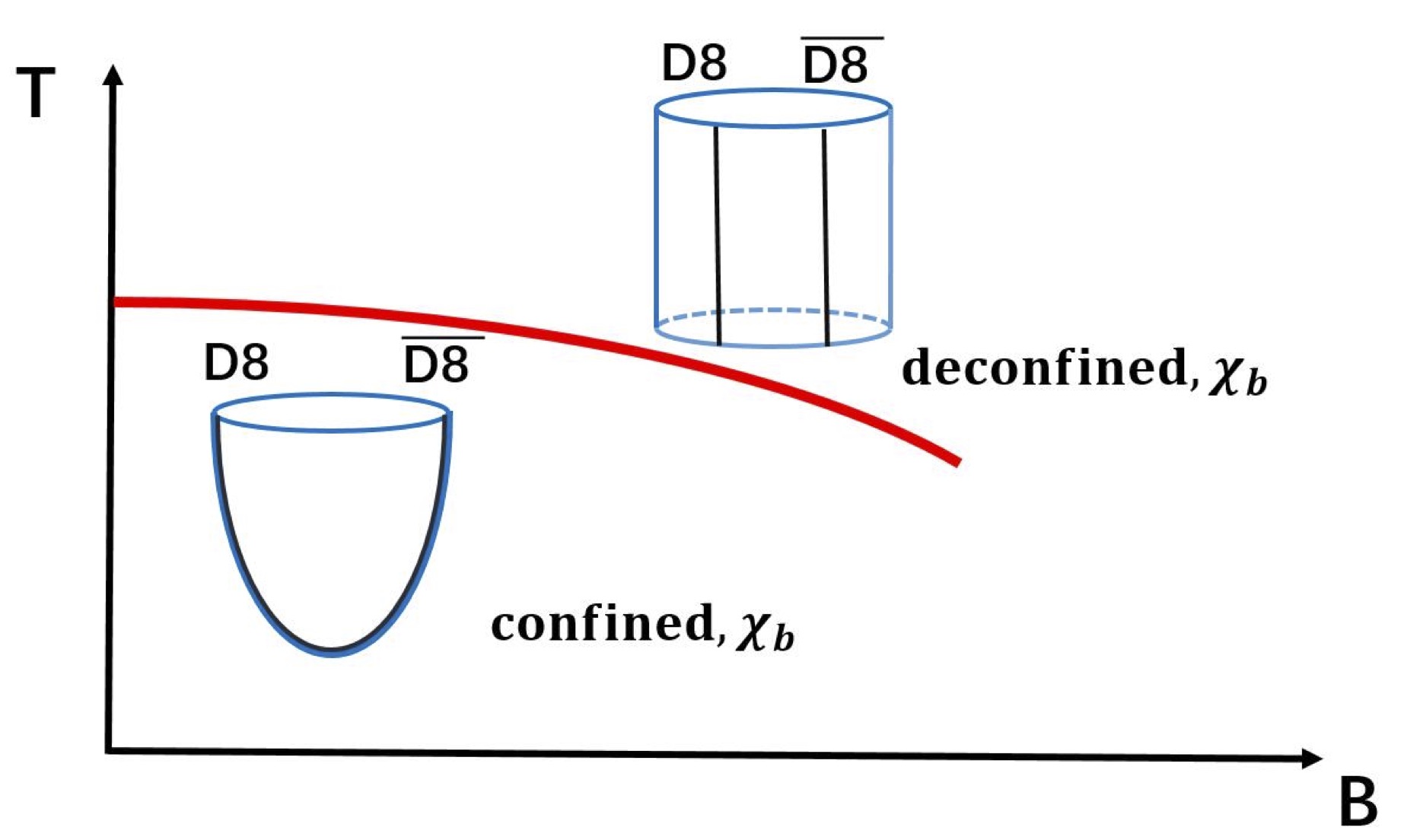}
\par\end{centering}

\begin{centering}
\includegraphics[scale=0.25]{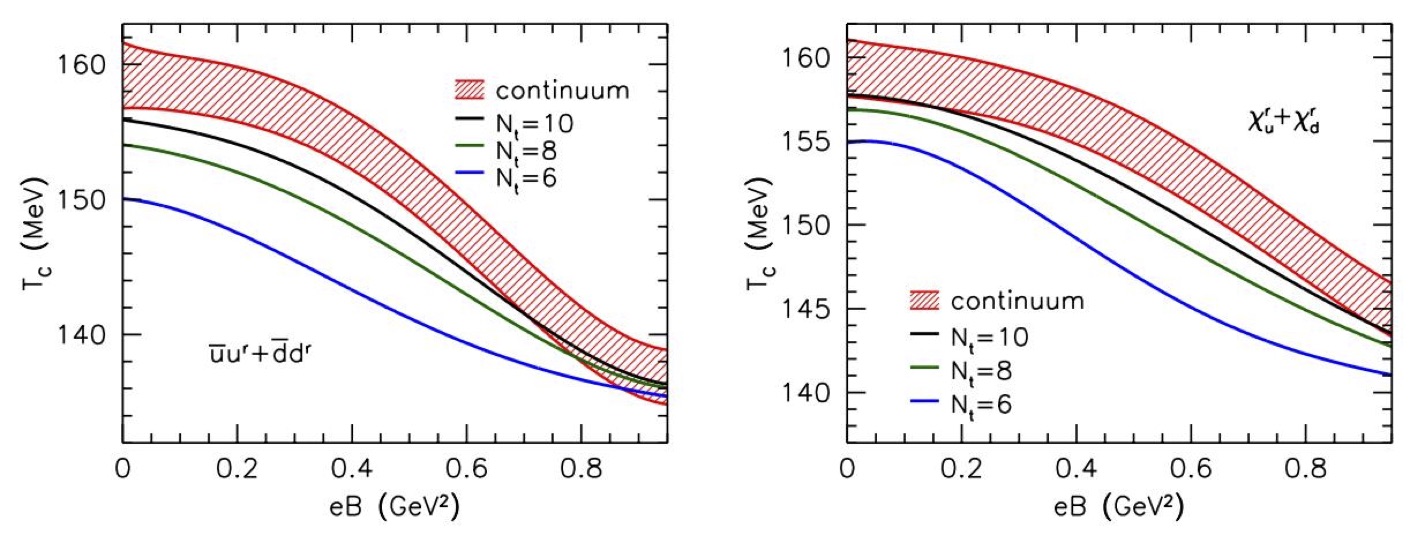}
\par\end{centering}

\caption{Holographic phase diagram V.S. lattice QCD result. \textbf{Upper}:
The phase diagram in T-B plan from our holographic formula (\ref{eq:85})
(\ref{eq:88}). \textbf{Lower}: The phase diagram in T-B plan from
some lattice QCD results in \cite{key-1}. $N_{t}$ is a parameter
in the lattice calculations.}
\end{figure}

\subsection{The backreaction case}

Let us turn to the case of backreaction. To get the phase diagram,
first we need to imposing (\ref{eq:53}) and (\ref{eq:54}) on (\ref{eq:69})
and (\ref{eq:73}), thus obtain the pressure of each phase as,

\begin{align}
p_{\mathrm{conf}.} & =\frac{2N_{c}^{2}\lambda_{4}M_{KK}^{4}}{3^{7}\pi^{2}}\bigg[1-\frac{\lambda_{4}^{2}}{12\pi^{3}}\frac{N_{f}}{N_{c}}\bigg(-5A_{1}+A_{2}-\frac{2\pi^{3/2}\left(9+7\sqrt{3}\pi+21\log\frac{27}{16}\right)}{35\Gamma\left(-\frac{5}{3}\right)\Gamma\left(\frac{1}{6}\right)}\nonumber \\
 & \ \ \ -\frac{1729\pi^{3/2}\left(72+\sqrt{3}\pi+3\log\frac{27}{16}\right)}{69984\Gamma\left(\frac{4}{3}\right)\Gamma\left(\frac{25}{6}\right)}q_{b}^{2}\bigg)\bigg],\nonumber \\
p_{\mathrm{deconf}.} & =\frac{128\pi^{4}N_{c}^{2}T^{6}\lambda_{4}}{2187M_{KK}^{2}}\left[1+\frac{\lambda_{4}^{2}}{12\pi^{3}}\frac{N_{f}}{N_{c}}\left(\frac{4}{7}-a_{1}-a_{2}+5a_{3}+\frac{14}{3}q_{b}^{2}\right)\right].\label{eq:87}
\end{align}
In order to obtain the critical temperature at the phase transition
point, we could solve the equation $p_{\mathrm{deconf.}}(T=T_{c})=p_{\mathrm{conf.}}$
as in the probe approximation. Then we find the following relation
between the critical temperature $T_{c}$ and the magnetic field $B$,

\begin{align}
\frac{2\pi T_{c}}{M_{KK}} & =1+\frac{N_{f}}{N_{c}}\left(X\lambda_{4}^{2}+\frac{B^{2}}{M_{KK}^{4}}Y\right),\label{eq:88}
\end{align}
where

\begin{align}
X & \simeq0.0005,\ Y\simeq-2.44,\label{eq:89}
\end{align}
and we have also imposed the special choice (\ref{eq:31}) (\ref{eq:52})
to (\ref{eq:88}). 

So similar to the case of the probe approximation, (\ref{eq:88})
shows the critical temperature increases by the effect of the flavors
without magnetic field. Notice that the factor in front of the magnetic
field is also negative which shows that the behavior of (\ref{eq:88})
is in agreement with the probe approximation. However, (\ref{eq:88})
should be a scheme-dependent statement thus it depends on the choices
of the appropriate interpretation (also the numbers of colors and
flavors). We have to keep this in mind, since we are less clear about
the full relations between the integration constants and the constant
magnetic field $B$, we therefore use the same ansatz as the most
simple choice as \cite{key-18} for the undetermined integration constants.
The behavior of $T_{c}$ with $B$ (\ref{eq:88}) would be sensitive
to the relations between the integration constants and the constant
magnetic field. 

Since the backreaction in our gravity solution is a perturbation,
we could additionally require the following relations in order to
omit the above ambiguities,

\begin{align}
 & -5A_{1}+A_{2}-\frac{2\pi^{3/2}\left(9+7\sqrt{3}\pi+21\log\frac{27}{16}\right)}{35\Gamma\left(-\frac{5}{3}\right)\Gamma\left(\frac{1}{6}\right)}-\frac{1729\pi^{3/2}\left(72+\sqrt{3}\pi+3\log\frac{27}{16}\right)}{69984\left(\frac{4}{3}\right)\Gamma\left(\frac{25}{6}\right)}q_{b}^{2}\nonumber \\
= & \ \ \frac{32\pi^{3/2}}{21\Gamma\left(\frac{1}{6}\right)\Gamma\left(\frac{4}{3}\right)}+\frac{4\pi^{3/2}}{3\Gamma\left(\frac{1}{6}\right)\Gamma\left(\frac{4}{3}\right)}q_{b}^{2},\label{eq:90}
\end{align}
and

\begin{equation}
-a_{1}-a_{2}+5a_{3}+\frac{14}{3}q_{b}^{2}=2q_{b}^{2},\label{eq:91}
\end{equation}
by comparing the thermodynamical quantities (\ref{eq:87}) with (\ref{eq:79})
(\ref{eq:85}) in the probe limit. It is consistent that our gravity
solution does not describe the full reactions from the flavors, because
all our calculations are in the Veneziano limit. On the other hand,
it is not necessary to discuss the full D4-$\mathrm{D}8/\overline{\mathrm{D}8}$
solution since it provides a totally different holographic duality
from the Witten-Sakai-Sugimoto model, which can not be described by
our perturbative solution in Section 3. In this sense, according to
(\ref{eq:90}) (\ref{eq:91}) with some special choices, all the integration
constants could be determined by the thermodynamical constraints.
However it implies that everything discussed in our manuscript can
not go beyond the probe approximation in fact.

\section{Summary and discussion}

In this paper, by considering the backreaction of the flavors and
the magnetic field, we have constructed gravitational solutions as
a magnetic-dependently holographic background in the Witten-Sakai-Sugimoto
model. Thus it corresponds to a large $N_{c}$ quantum field theory
(or large $N_{c}$ QCD) with dynamical flavors in an external magnetic
field. We have proved out our gravitational solutions satisfy their
equations of motion explicitly in the first order of $N_{f}/N_{c}$.
The solutions are analytic both in confined (bubble) and deconfined
(black brane) case at low (zero) or high (finite) temperature. Therefore
these solutions are able to study the the influence of dynamical flavors
in an external magnetic field as a holographic version of \cite{key-1,key-2,key-3,key-4}.
In order to determine the integration constants in our solutions,
we require the backgrounds are completely regular in the IR region
of the dual field theory as the unflavored case since the flavors
are small perturbations. On the other hand, we also try to turn off
the sources or VEVs of some gauge invariant operators in the dual
field theory as another constraint. However the calculation shows
it is not enough to determine all the integration constants just by
these two constraints. So we have to keep those undetermined integration
constants as some generic parameters temporarily. 

In order to compare our magnetic-dependent case with \cite{key-18},
we simply chose the same value for the undetermined integration constants
as \cite{key-18}, to study some physical properties about hadronic
physics in an external magnetic field, such as the running coupling,
(QCD) string tension, baryon mass. We find the UV behavior of the
running coupling is not affected by the presence of the magnetic field.
And the string tension, the mass of baryon increase by the presence
of the flavor or the magnetic field. But we need to keep in mind these
behaviors should depend on which scheme is chosen and where some observables
is kept fixed in the theory since theories with different numbers
of flavors might be different. Additionally, due to the simply choice
as \cite{key-18} for those undetermined integration constants, the
results (in this part) are not strictly rigorous thus some of them
might still seem unrealistic. 

Moreover, it shows the physical significance of our work by investigating
the holographic renormalization and thermodynamics with our magnetic-dependently
gravitational solution. We employ the counterterm \cite{key-29} and
its covariant formula \cite{key-18} for this model then evaluate
them by our magnetic solution. The motivation for studying this counterterm
is to renormalize the free energy, to study the Hawking-Page transition
holographically in the presence of the magnetic field. In some applications
of the Witten-Sakai-Sugimoto model, holographic renormalization may
not be necessary for studying the phase transition. Since those concerns
are the difference of the free energy of the various configurations
of the flavor branes in the same background, which is not the Hawking-Page
transition of this model. So the difference of the free energy could
be finite in those approaches (such as \cite{key-11,key-15,key-19,key-16,key-30}).
However, in our calculations, holographic renormalization is needed
since we (more than that) also consider the transition between differently
geometric background. According to our calculations, if the parameters
in the covariant counterterms are allowed to depend on the magnetic
field as \cite{key-29}, we find the present counterterms are enough
to cancel all the divergences.

In particular, after the holographic renormalization, we have concentrated
the attention on the holographic phase diagrams in the presence of
the magnetic field, and compare it with lattice QCD results. In our
backreaction case, we find the pressure of both phases evaluated by
our magnetic-dependent solution agrees with \cite{key-29} qualitatively.
Although the behavior of the phase diagram agrees with lattice QCD
\cite{key-1,key-2}, there might be a bit ambiguous since we have
chosen the special value for the integration constants. In the probe
approximation, the phase diagram is clear and also in agreement with
the lattice QCD \cite{key-1,key-2} qualitatively (Figure 2). Thus
it could be interpreted as the inhibition of confinement or chirally
broken symmetry by the magnetic field holographically. Besides, we
additionally require our backreaction solution coincides with the
case of probe limit by the analyses of the thermodynamics, so that
all the integration constants could be determined in this sense.

Finally, let us comment something more about our work. As an improvement
to \cite{key-29}, we have employed the technique used in \cite{key-18}
to take into account the backreaction from flavors and the magnetic
field. Because of the presence of the magnetic field, actually we
need to solve a set of highly non-linear equations of motion first
to obtain a magnetic-dependently gravitational solution, as shown
in (\ref{eq:23}) and (\ref{eq:43}). Since it is hopeless to find
an analytic solution from these extremely complicated equations, we
solve them by keeping the leading $B^{2}$ terms. So while it is a
challenge to keep all the orders of the DBI action to solve analytically,
some numerical calculations might be worthy. Besides, during our calculations,
we have restricted that $\mathrm{D}8/\overline{\mathrm{D}8}$-branes
are placed at antipodal points of $x_{4}$ - circle in the confined
phase. So to extend this part to the non-antipodal case would be natural,
and the chiral symmetry could also be restored after deconfinement
transition. Moreover, it is also interesting to turn on a chemical
potential and a magnetic field together on the flavor branes in this
framework, since a similar phenomena, named as ``inverse magnetic
catalysis'', has also been found by using this model in the probe
approach of \cite{key-19}. However, there would be a non-vanished
Chern-Simons term necessarily\footnote{The Witten-Sakai-Sugimoto model would be similar to the Einstein-Maxwell
system if considering the bulk field and expanded DBI action by small
$F$ ($U\left(1\right)$ gauge field strength). There have been some
discussions about the \textquotedblleft inverse magnetic catalysis\textquotedblright{}
in the Einstein-Maxwell system as \cite{key-19,key-45}. However,
as a difference from Einstein-Maxwell system and also a computational
challenge, we have to consider the additional Romand-Romand field
in the bulk and the non-vanished Chern-Simons (or Wess-Zumino) term
if taking into account the backreaction from the flavor branes (full
action). While the computation is difficult, it would be quite interesting
for a future study.} if turning on the chemical potential and the magnetic field together
as \cite{key-19,key-30}. It would be more difficult to search for
an analytic solution even in the expansion of small baryon charge,
magnetic field and $N_{f}/N_{c}$ in that case since the equations
of motion would be complicatedly coupled to each other once the backreaction
is considered. We would like to leave these interesting topics for
a future study to improve our calculations about holographic QCD.

\section*{Acknowledgments}

This work is inspired by our previous work \cite{key-16} in USTC,
and also by \cite{key-46} the recent research on the magnetic field
in heavy-ion collision from our colleagues. And we would like to thank
Andreas Schmitt, Prof. Qun Wang and Dr. Chao Wu for helpful discussions.

\end{document}